%% file: main.tex
\documentclass[journal]{IEEEtran}
\usepackage{amsmath,amsfonts}
\usepackage{algorithm}
\usepackage{array}
\usepackage[caption=false,font=normalsize,labelfont=sf,textfont=sf]{subfig}
\usepackage{textcomp}
\usepackage{stfloats}
\usepackage{url}
\usepackage{verbatim}
\usepackage{rotating}
\usepackage{braket}
\usepackage{algorithm,algpseudocode}
\usepackage{tikz-qtree}
\usepackage{mathtools}
\usetikzlibrary{positioning}
\usepackage{graphicx}
\usepackage{tikz}
\usepackage{cite}
\usepackage[qm]{qcircuit}
\begin{document}
% was: Quantum Optimization for Computational Intelligence: An Introduction with Tutorials
% jay: How Quantum Optimization can be used for Computational Intelligence?
% Quantum Approximate Optimization as a Computational Intelligence Technique
\title{Quantum Approximate Optimization: A Computational Intelligence Perspective}
%\title{\christo{Quantum Ansatz for Computational Intelligence: A Combinatorial Optimization Perspective}}
%\title{\christo{Quantum Alternating Operator Ansatz:  A Combinatorial Optimization Framework}}
\author{Christo Meriwether Keller, New Mexico State University, USA\\ Satyajayant Misra, New Mexico State University, USA \\
Andreas B\"artschi, Los Alamos National Laboratory, USA \\
Stephan Eidenbenz, Los Alamos National Laboratory, USA \\
%$^1$New Mexico State University, USA; $^2$Los Alamos National Laboratory, USA}
}
\markboth{-}%
{Shell \MakeLowercase{\textit{et al.}}: A Sample Article Using IEEEtran.cls for IEEE Journals}

\IEEEpubid{ }

\maketitle

\begin{abstract}
\textcolor{black}{Quantum computing is an emerging field on the multidisciplinary interface between physics, engineering, and computer science with the potential to make a large impact on computational intelligence (CI).  
The aim of this paper is to introduce quantum approximate optimization methods to the CI community because of direct relevance to solving combinatorial problems. 
We introduce quantum computing and  variational quantum algorithms (VQAs). VQAs are an effective method for the near-term implementation of quantum solutions on noisy intermediate-scale quantum (NISQ) devices with less reliable qubits and early-stage error correction. Then, we explain Farhi et al.'s quantum approximate optimization algorithm (Farhi's QAOA, to prevent confusion). This VQA is generalized by Hadfield et al. to the quantum alternating operator ansatz (QAOA), which is a nature-inspired (particularly, adiabatic) quantum metaheuristic for approximately solving combinatorial optimization problems on gate-based quantum computers.  %Based on a discretization of an adiabatic process, QAOA imitates simulated annealing on NISQ devices.
We discuss connections of QAOA to relevant domains, such as computational learning theory and genetic algorithms, discussing current techniques and known results regarding hybrid quantum-classical intelligence systems.  We present a schematic of how QAOA is constructed, and also discuss how CI techniques can be used to improve QAOA.
We conclude with QAOA implementations for the well-known maximum cut, maximum bisection, and traveling salesperson problems, which can serve as templates for CI practitioners interested in using QAOA.} 
\end{abstract}


\section{Introduction}
\IEEEPARstart{Q}{uantum computing} is an emerging technology promising to disrupt various fields of industry.  For example, Shor's polynomial-time quantum algorithm for the factoring problem has spurred an ongoing search for robust `post-quantum' cryptosystems.  However, present-day quantum computers are noisy and their performance tends to decrease sharply with increases in scale.  General purpose `noisy intermediate-scale quantum' (NISQ) computers are unable to reliably factor semiprimes as small as $35=5\times 7$ with Shor's algorithm, only succeeding around 14\% of the time \cite{shorexperimental}.

%Better, semiprimes as large as $376289=659\times 571$ have been factored using quantum annealers, although to factor a semiprime $N$ with this method requires $\mathcal{O}(\ln(N)^2)$ reliable qubits~\cite{annealerexperimental}.  
Shor's algorithm can be implemented on universal gate-based quantum computers with as few as $\mathcal{O}(\ln(N))$ qubits~\cite{shorqubits}.  Thus breaking current versions of RSA would take far more reliable qubits than we currently have, 
%no matter what the approach.   
laying out of reach of NISQ devices---this is generally true for the use of most promising quantum algorithms. This \textit{status quo} has generated interest in variational quantum algorithms (VQAs), which have emerged as the `leading proposal for achieving quantum advantage on near-term quantum computers'~\cite{vqas}.  For example, the variational quantum eigensolver (VQE)  has been used for quantum chemical simulations of molecules on NISQ devices~\cite{h12,co2o2}. These hybrid quantum-classical algorithms take a learning-based approach, parameterizing a quantum `ansatz' and classically training it to give better results.  Thus error-prone quantum overhead is offloaded into classical computation and better results achieved without giving up the potential for quantum advantanges.

The quantum approximate optimization algorithm (QAOA), proposed by Farhi et al.~\cite{qaoafarhi}, is a promising VQA metaheuristic for combinatorial optimization. To clarify their derivation of QAOA, we first describe the quantum adiabatic algorithm (QAA)~\cite{qaa}, a strategy comparable to simulated annealing but for \textit{quantum} annealers (or, hypothetically, adiabatic quantum computers).  Discretization of QAA results in QAOA, a heuristic approximate unconstrained optimization VQA suitable for general purpose quantum computers. QAOA
%{\color{green}{ framework/algorithm}} %% Farhi's QAOA is NOT a framework yet, only an algorithm
was further generalized by Hadfield et al.~\cite{hadfield2019} into a \emph{broader framework} 
%% emphasizing that Hadfield's QAOA is now a framework
for both constrained and unconstrained optimization on near-term quantum computers, resulting in what they termed the quantum alternating operator ansatz (with the same abbreviation, QAOA).   \textcolor{black}{That is, Farhi's QAOA algorithm inspired Hadfield's framework, which it is a special case of. The term QAOA is ambiguous, with QAOA  referring to both Farhi's original algorithm and Hadfield's framework in the quantum computing literature. 
%% Added by Andreas, I believe this is important to mention this way
In the rest of this paper, QAOA refers to Hadfield's framework. To refer to the original algorithm, we use the term Farhi's QAOA.  
%This distinction is mostly restricted to the sections discussing the history and derivation of QAOA.
}

This ansatz is nature-inspired in the sense that it is comparable to simulated annealing, and adiabatic in how it approximates the adiabatic evolution of a physical system.  QAOA has been applied to many important real-world problems.  It has promising applications to decision problems in quantitative finance, for instance to discrete portfolio optimization~\cite{fin_matwiejew} and constrained transaction settlement~\cite{transaction_settlement_qaoa}.  Also, it has been applied to more traditional engineering problems such as vehicle routing~\cite{vehicleqaoa} and wireless network scheduling~\cite{networkqaoa}. \IEEEpubidadjcol

Moreover, in its full generality, QAOA can be applied to nearly any combinatorial optimization problem. This makes it a metaheuristic comparable to other computational intelligence techniques such as genetic algorithms, ant colony optimization, and particle swarm optimization.  Conversely, techniques such as genetic algorithms and reinforcement learning can be applied to QAOA itself, in large part because they help avoid suboptimal local optima in complex search spaces.  We will discuss comparisons to and connections with these topics, and present potential future work in such directions. 

The advantages of these techniques come when problems have complicated solution spaces: for example, when these spaces are constrained or not convex.  In these situations simple methods like classical gradient descent can get stuck in local minima, or exponentially slowed down at saddle points~\cite{gradient_saddle}. QAOA has a theoretical guarantee to converge to an optimal solution provided a reliable quantum computer, long runtimes, and properly chosen parameters, suggesting the potential for advantage.  Each of these areas is a current topic of research; for example, Wang et al.~\cite{wang2018quantum} provides an analytic expression for optimal level-1 QAOA parameters. In practice the parameters are numerically optimized (e.g. by gradient descent), and the convergence is not always realized.  Therefore, the practical benefits of QAOA should come from heuristically leveraging superpositions of solutions to find good approximate solutions more quickly than classically possible.

In Section~\ref{sec2}, we introduce critical background information: formalisms for combinatorial optimization problems and quantum computing.  We introduce two models, adiabatic computing and the unitary gate model, which are relevant to our subject.  A discussion of variational quantum algorithms follows, then some remarks on combinatorial metaheuristics. 

In Section~\ref{sec3} we describe three algorithms.  First, we look at the quantum adiabatic algorithm (QAA) to exactly solve combinatorial optimization problems on quantum annealers, which is then adapted as an approximation algorithm for general-purpose quantum computers called the quantum approximate optimization algorithm (Farhi's QAOA).  Then we discuss the generalization of Farhi's QAOA to a robust metaheuristic called the quantum alternating operator ansatz (QAOA).  We look at recent directions and extensions of this ansatz: modified QAOA ans\"atze%
% (plural of \textit{ansatz})
, empirical studies, and analytic results.  % \stephan{Not sure this discussion is necessary re German, but if you leave it you perhaps also have to say: in keeping with academic tradition we do not capitalize it as would be required in proper German.}

Section~\ref{sec4} connects QAOA to computational intelligence more broadly, in particular to ant colony optimization, genetic algorithms, simulated annealing and reinforcement learning.  These techniques can be applied to QAOA, and conversely QAOA can be applied to them. This suggests future work in both directions with other CI techniques.

Section~\ref{sec5} explores the process of QAOA algorithm design.  We discuss the mapping of QAOA parameters to the maximum cut problem and the asymmetric traveling salesperson problem, showing implementations of crucial components in Qiskit~\cite{Qiskit}.  We modify the maximum cut instance to deal with the constrained subspace search problem known as maximum bisection. This section provides useful examples to practitioners looking to design QAOA instances for their own problems.

Section~\ref{sec6} concludes with a high-level review of the paper and discussion of further ideas for work in these directions.

\section{Background}
\label{sec2}
Before getting into the details of the quantum algorithms, we explain fundamental background topics: combinatorial optimization problems, models for quantum computing, and a formalism for variational quantum algorithms.

\subsection{Combinatorial Optimization}

Optimization problems are generally divided into two groups: those with continuous variables and those with discrete variables~\cite{papasteig}.  Some problems, such as some with sigmoid utility, can be considered to have both types of variables~\cite{sigmoid}, although they are usually treated as continuous.  In this paper we focus on problems with purely discrete variables, which we call `combinatorial'. Combinatorial optimization problems can be formalized in different ways, and here we present a formalism that is convenient for constructing our algorithms.

\underline{Definition.}  An \textit{instance} for or \textit{model} of a combinatorial optimization problem is a pair $(S,f)$  such that
\begin{itemize}
    \item $S$ is a finite set of $n$-bit strings called the \textit{feasible solution space} or \textit{search space} for the problem.
    \item $f$ is a function %$\mathbf{2}^n\to \mathbb{R}$ c
                $S \to \mathbb{R}$ called the \textit{cost function}, with respect to which we are to optimize.
\end{itemize}

A solution to such an instance $(S,f)$ is a bitstring $z\in S$ such that $z=\max_S f$. Of course, a minimization problem for the same situation can be constructed by taking $-f$ as the cost function instead.  In this sense, we say that maximization and minimization are dual to one another.  Also, we say a problem is \textit{unconstrained} if $S=\mathbf{2}^n$ and $\textit{constrained}$ otherwise.
\IEEEpubidadjcol

Combinatorial optimization problems are important in computer science, where they provide  fundamental examples of NP-hard problems such as the traveling salesperson problem and maximum satisfiability.  In general, the decision versions of NP-hard problems belong to the NP-complete class.
 Many  examples are collected in the classic book ``Computers and Intractability'' by Garey and Johnson~\cite{gareyjohnson}.  %As a historical note, the techniques in this paper were first applied to the maximum cut (MaxCut) problem.  MaxCut is dual to the minimum cut (MinCut) problem, \christo{the decision version of which} was one of 21 early examples of NP-complete problems given by Richard Karp in his seminal paper on NP-completeness~\cite{maxcut}.\stephan{I would quote Garey and Johnson in addition or instead of Ausiello. Note that an optimization problem formally is not in NP, because NP is only decision problems, you can get around this by speaking of NP-hard problems (not NP problems)}

\IEEEpubidadjcol 

\subsection{Linear Algebraic Concepts}
In the rest of this section, we assume a basic familiarity with linear algebra and quantum computing (for a refresher, we suggest  Chapters 1-4 of Nielsen and Chuang~\cite{nielsenchuang} and Abhijith et al.~\cite{abhijith2022quantumalgorithms}).  However, we recall the most important concepts here to keep this work relatively self-contained. We note that the mathematical details in what follows are presented for completeness; an exact understanding is not necessary for understanding the overall concept(s).

 \underline{Definition.} An $n$-qubit \textit{quantum state} $\ket{\psi}$  is a nonzero element of the vector space $V=\mathbf{C}^{2^n}$ with the equivalence relation $\sim$ that for $\ket{\psi},\ket{\phi}\in V$, $\ket{\psi}\sim\ket{\phi}$ iff $\ket{\phi}=\lambda\ket{\psi}$ for some nonzero scalar $\lambda\in\mathbf{C}$.  For simplicity we take normalized vectors as class representatives. Briefly then, a state is an element of $\mathbf{CP}^{2^n-1}$, the complex projective space of $\mathbf{C}^{2^n}.$
 %\stephan{Not everyone understands CP, what is $\phi$ supposed to be?}

For a one-qubit system, we take the \textit{standard} basis vectors $\ket{0}$ and $\ket{1}$ as primitives, and write vectors using the Dirac bra-ket notation that vectors $\left(\begin{smallmatrix}
  \alpha\\\beta\end{smallmatrix}\right)\in V$ are \textit{kets}
  $\ket{v}=\alpha\ket{0}+\beta\ket{1}$ and  $\left(\begin{smallmatrix} 
  \overline{\alpha} & \overline{\beta}  \end{smallmatrix}\right)$ of vectors 
  are \textit{bras} $\bra{v}:=\ket{v}^\dagger$.  This gives the notational 
  nicety that the inner product of $u$ and $v$ can be written as a \textit{bra-ket}
  $\braket{u|v}$.  \textit{Measuring} a one qubit state
  $\alpha\ket{0}+\beta\ket{1}$ returns $\ket{0}$ with probability $||\alpha||^2$ and $\ket{1}$ with probability $||\beta||^2$. This extends to tensor product states in the obvious way, and corresponds to the \textit{wave-function collapse} of a quantum state as defined in quantum mechanics.  For example, the expectation value (weighted average of outcomes) of an $n$ by $n$ operator $M$ on a state $\ket{\psi}\in V^n$ is given by $\braket{\psi|M|\psi}$.

For example, the one-qubit \textit{plus state} representing the equal superposition of compuational basis states 
\[\ket{+}=\frac{1}{\sqrt{2}}\begin{pmatrix}1\\1\end{pmatrix}=\frac{1}{\sqrt{2}}\ket{0}+\frac{1}{\sqrt{2}}\ket{1}\] is equally likely to be measured in either state $\ket{0}$ or $\ket{1}$, and these are the only options.  That is, $\braket{+|0}^2=\braket{+|1}^2=1/2$.  This state will be useful later, as its tensor product $\ket{+}^{\otimes n}$ encodes the unconstrained search space $\mathbf{2}^n$ as a quantum state. 

We will deal with two classes of matrices in this paper, which we call unitary gates and Hermitian operators:
\begin{itemize}
\item A \textit{unitary gate} is a complex square matrix $M$ such that $M^{-1}=M^\dagger$.  That is, $M$ is unitary if its inverse is its conjugate.  These operations are important in quantum mechanics since they preserve inner products.%probability amplitudes.

\item A \textit{Hermitian operator} is a complex square matrix $M$ such that $M^{\dagger}=M$.  That is, $M$ is Hermitian if it is its own conjugate.  These operations are important in quantum mechanics since they encode the energy of a system, and every unitary gate can be written as a matrix exponential of some Hermitian scaled by $i$ (unit time evolution).
\end{itemize}

For some important examples, we have the \textit{Pauli matrices} 
\[X=\begin{pmatrix}0 & 1\\ 1&0\end{pmatrix},\ Y=\begin{pmatrix}0 & -i\\ i & 0\end{pmatrix},\ Z=\begin{pmatrix} 1 & 0\\ 0 & -1\end{pmatrix}\]

\noindent which are clearly unitary and Hermitian, and can be used in quantum algorithms.  Notably, the Pauli-X gate generalizes the classical NOT gate.  Additionally, because these are Hermitian, the matrix exponents of their tensor products are all unitary as well. Also relevant to us is the unitary \textit{Hadamard gate}
\[H=\frac{1}{\sqrt{2}}\begin{pmatrix}1 & 1\\ 1& -1\end{pmatrix}\]
which sends $\ket{0}$ to $\ket{+}$ and $\ket{1}$ to $\ket{-}=_{df}\frac{1}{\sqrt{2}}\ket{0}-\frac{1}{\sqrt{2}}\ket{1}$, and serves as a change of basis operator between the \textit{computational basis} $\{\ket{0},\ket{1}\}$ and the \textit{Hadamard basis} $\{\ket{+},\ket{-}\}$.

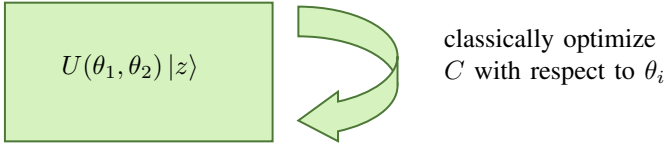
\begin{figure}
\tikzset{every picture/.style={line width=0.75pt}} %set default line width to 0.75pt        
\begin{tikzpicture}[x=0.75pt,y=0.75pt,yscale=-1,xscale=1]

%Shape: Rectangle [id:dp2301664466483393] 
\draw  [color={rgb, 255:red, 76; green, 120; blue, 54 }  ,draw opacity=1 ][fill={rgb, 255:red, 214; green, 242; blue, 190 }  ,fill opacity=1 ] (143,104) -- (278,104) -- (278,174) -- (143,174) -- cycle ;
%Curve Right Arrow [id:dp9850613520885625] 
\draw  [color={rgb, 255:red, 76; green, 120; blue, 54 }  ,draw opacity=1 ][fill={rgb, 255:red, 214; green, 242; blue, 190 }  ,fill opacity=1 ] (341,146) .. controls (341,132.19) and (318.61,121) .. (291,121) -- (291,106) .. controls (318.61,106) and (341,117.19) .. (341,131) ;\draw  [color={rgb, 255:red, 76; green, 120; blue, 54 }  ,draw opacity=1 ][fill={rgb, 255:red, 214; green, 242; blue, 190 }  ,fill opacity=1 ] (341,131) .. controls (341,141.25) and (328.66,150.06) .. (311,153.92) -- (311,148.92) -- (291,163.5) -- (311,173.92) -- (311,168.92) .. controls (328.66,165.06) and (341,156.25) .. (341,146)(341,131) -- (341,146) ;

% Text Node
\draw (363,116) node [anchor=north west][inner sep=0.75pt]   [align=left] {classically optimize \\$\displaystyle C$ with respect to $\displaystyle \theta _{i}$};
% Text Node
\draw (170,128.4) node [anchor=north west][inner sep=0.75pt]    {$U( \theta _{1} ,\theta _{2})\ket{z}$};
\end{tikzpicture}
 \caption{A VQA with two variational classical parameters, optimizing $C\circ U\ket{z}$ with respect to $\theta_1$ and $\theta_2$, where $C$ is the cost function and $U$ is the ansatz.}

\end{figure}

Quantum features such as superposition, entanglement, and relative phase allow for much of the power of quantum computing.  With this mathematical setup for our physical systems, we can now define our models of quantum computing.

\subsection{Models for Quantum Computing}

In this paper we will work with two models for quantum computing: the adiabatic model for quantum annealers, and the unitary gate model.  These models are polynomially-equivalent, which means that problems have essentially the same complexity no matter which is used~\cite{adiabaticmodel}.  However, the gate-based model is preferred for most real-world applications.

\begin{enumerate}
    \item \textit{Adiabatic computing.} In this model, an algorithm is defined by two Hermitian matrices $H_0$ and $H_1$.  We initialize our quantum annealer to a ground state of $H_0$ and slowly evolve the system along the adiabatic path $H(t)=(1-t)H_0+tH_1$, where $t$ is the time elapsed in the adiabatic computation.  The adiabatic theorem guarantees that if we start in such a state of $H_0$, we end up in a ground state of $H_1$ after such evolution. The complexity of an adiabatic algorithm is the actual amount of time that it takes to run, which is on the order of the inverse of the minimum spectral gap $g_{min}$ of the Hermitian matrices along the adiabatic path \cite{adiabaticmodel}.
    
    %The runtime of the algorithm is given by the actual time it takes to run, which depends on the minimal spectral gap of $H(t)$ between $t=0$ and $t=1$. Additionally, $H_0$ and $H_1$ must be $k$-local on $d$-states with unique ground states and the ground state for $H_0$ should be a tensor product state. \cite{adiabaticmodel}
    \item \textit{Unitary gate model.}  We begin in a standard basis state and apply finitely many unitary matrices called \textit{elementary gates}. This procedure puts us in another state, which we measure. We use unitaries because they exactly preserve the norm.  %For example, the zero matrix would take us out of our projective space altogether.
    Additionally, we define the (gate) complexity of a unitary gate algorithm as the number of elementary gates used.
\end{enumerate}

There are restrictions on the choices of $H_0$ and $H_1$ for adiabatic computing which are beyond the scope of this paper.  We will only briefly discuss the quantum adiabatic algorithm before approximating it by a variational quantum algorithm. 

\subsection{Variational Quantum Algorithms}

A variational quantum algorithm (VQA) is a hybrid quantum-classical approach that consists of three parts: ({\it i})~a parameterized quantum circuit called an \textit{ansatz}, ({\it ii})~a cost function rating measurements of the quantum circuit, and ({\it iii})~a classical optimization method or learning algorithm.  A classical outer loop improves the ansatz parameters by computing the cost function on the output.  This allows for a reduction in the depth of quantum circuits, enabling better results on NISQ devices.  The model of VQAs actually allows for universal computing, but this assumes robust error correction~\cite{univ_vqa}.

Recently, VQAs have been applied to many important computational problems.  For example, the \textit{variational quantum eigensolver} (VQE) is a VQA for quantum chemical simulations.  The VQE and its extensions have been applied to simulate simple molecules such as lithium hydride, carbon monoxide, carbon dioxide, and hydrogen chains as long as dodecahydrogen~\cite{h12,co2o2}.  See Fig. 1 for an abstract representation of the VQA model with two variational parameters, where $U$ is the quantum ansatz, $\theta_1,\theta_2$ are two variational parameters, and $C$ is the cost function.  The arrow represents an unspecified optimization algorithm with some convergence condition.

However, VQAs like VQE come with a downside related to complexity theory.  Finding the ground states of even very restricted Hamiltonians belongs to a class of problems called QMA-complete, which is ``the quantum analogue of NP in a probabilistic setting''~\cite{hamiltonians}. Then if VQE solves every instance of the Hamiltonian ground state problem, it is in BQP (the quantum equivalent of P in a probabilistic setting) so that $\textnormal{QMA}\subseteq \textnormal{BQP}$. Now by analogy with P and NP this seems unlikely.
Even for optimization problems in $\textnormal{NP} \subseteq \textnormal{QMA}$, it is commonly believed that $\textnormal{NP} \not \subseteq \textnormal{BQP}$. Furthermore, optimizing the parameters for QAOA is a non-convex problem and may exhibit barren plateaus~\cite{cerezo2021cost}, 
so we expect general VQA techniques such as QAOA and VQE to remain an approximate heuristic for most problems. Still, there is reason to believe quantum advantage can be achieved through these means.  Recent numerical work by Boulebnane and Montanaro~\cite{boulebnane2022solving}, Golden et al.~\cite{golden2023numerical} and Shaydulin et al.~\cite{shaydulin2023evidence}
suggest advantages for QAOA in certain domains, although a similar claim was once defeated by a new classical algorithm~\cite{farhi2015quantum,barak2015beating}.

Many applications of VQAs have been studied and surveyed~\cite{vqas}.  Most notable for us is combinatorial optimization, where QAOA is the leading VQA.  VQAs themselves, and QAOA in particular, can additionally be viewed as application areas for computational intelligence techniques.  For example, when optimal ansatz parameters are not known, techniques like reinforcement learning have been used to achieve better results~\cite{reinforcementqaoa,qaoareinforcement}, and genetic algorithms are a natural choice for compiling QAOA circuits~\cite{geneticcompile,geneticqcc}.

\subsection{Combinatorial Metaheuristics}

Metaheuristics are 
`problem agnostic techniques that establish an iterative search process to find an optimal solution for a given problem'~\cite{metaheuristics}.  Notable combinatorial metaheuristics  include ant-based optimization~\cite{ants} and evolutionary transfer optimization \cite{eto}.  Over two hundred optimization metaheuristics are collected in a paper by Ezugwu et al.~\cite{metas}.

Roughly, a metaheuristic is a parameterized algorithm where parameters are chosen, optimized, or iteratively constructed to fit a particular problem.  In this way VQAs as a whole are metaheuristics insofar as their quantum parts are parameterized and classically optimized.  QAOA is a metaheuristic: it is defined by general parameters (both classical and quantum) that can handle different classes of optimization problems.  

Metaheuristics are often used to approximately solve problems for which no efficient deterministic method is known.  These problems often have non-convex search spaces or search spaces superpolynomial in the input size.  Many important NP-complete problems fall into these categories.

\begin{figure*}
    \centering
    \tikzset{every picture/.style={line width=0.75pt}} %set default line width to 0.75pt        

\begin{tikzpicture}[x=0.75pt,y=0.75pt,yscale=-1,xscale=1]

%Straight Lines [id:da22528067311212796] 
\draw [color={rgb, 255:red, 76; green, 120; blue, 54 }  ,draw opacity=1 ][fill={rgb, 255:red, 214; green, 242; blue, 190 }  ,fill opacity=1 ]   (21,117.67) -- (52,117.67) ;
%Straight Lines [id:da9446647413192386] 
\draw [color={rgb, 255:red, 76; green, 120; blue, 54 }  ,draw opacity=1 ][fill={rgb, 255:red, 214; green, 242; blue, 190 }  ,fill opacity=1 ]   (21,157.67) -- (51,157.67) ;
%Shape: Rectangle [id:dp2041100220624501] 
\draw  [color={rgb, 255:red, 76; green, 120; blue, 54 }  ,draw opacity=1 ][fill={rgb, 255:red, 214; green, 242; blue, 190 }  ,fill opacity=1 ]  (52,107.67) -- (112,107.67) -- (112,166.67) -- (52,166.67) -- cycle ;
%Straight Lines [id:da5584303350728332] 
\draw  [color={rgb, 255:red, 76; green, 120; blue, 54 }  ,draw opacity=1 ][fill={rgb, 255:red, 214; green, 242; blue, 190 }  ,fill opacity=1 ]  (113,118.67) -- (144,118.67) ;
%Straight Lines [id:da40668721267697694] 
\draw  [color={rgb, 255:red, 76; green, 120; blue, 54 }  ,draw opacity=1 ][fill={rgb, 255:red, 214; green, 242; blue, 190 }  ,fill opacity=1 ]  (113,158.67) -- (143,158.67) ;
%Shape: Rectangle [id:dp18065061971939755] 
\draw [color={rgb, 255:red, 56; green, 160; blue, 216 }  ,draw opacity=1 ][fill={rgb, 255:red, 183; green, 230; blue, 255 }  ,fill opacity=1 ]  (144,108.67) -- (204,108.67) -- (204,167.67) -- (144,167.67) -- cycle ;
%Straight Lines [id:da15392480213618565] 
\draw [color={rgb, 255:red, 76; green, 120; blue, 54 }  ,draw opacity=1 ][fill={rgb, 255:red, 214; green, 242; blue, 190 }  ,fill opacity=1 ]   (205,119.67) -- (236,119.67) ;
%Straight Lines [id:da9262315254177915] 
\draw   [color={rgb, 255:red, 76; green, 120; blue, 54 }  ,draw opacity=1 ][fill={rgb, 255:red, 214; green, 242; blue, 190 }  ,fill opacity=1 ] (205,159.67) -- (235,159.67) ;
%Shape: Rectangle [id:dp0928579716614486] 
\draw  [color={rgb, 255:red, 148; green, 134; blue, 214 }  ,draw opacity=1 ][fill={rgb, 255:red, 216; green, 209; blue, 255 }  ,fill opacity=1 ] (236,109.67) -- (296,109.67) -- (296,168.67) -- (236,168.67) -- cycle ;
%Straight Lines [id:da048330393188396714] 
\draw  [color={rgb, 255:red, 76; green, 120; blue, 54 }  ,draw opacity=1 ][fill={rgb, 255:red, 214; green, 242; blue, 190 }  ,fill opacity=1 ]  (296,117.67) -- (327,117.67) ;
%Straight Lines [id:da21812976650155824] 
\draw   [color={rgb, 255:red, 76; green, 120; blue, 54 }  ,draw opacity=1 ][fill={rgb, 255:red, 214; green, 242; blue, 190 }  ,fill opacity=1 ] (296,157.67) -- (326,157.67) ;
%Straight Lines [id:da6949188181915573] 
\draw  [color={rgb, 255:red, 76; green, 120; blue, 54 }  ,draw opacity=1 ][fill={rgb, 255:red, 214; green, 242; blue, 190 }  ,fill opacity=1 ]  (364,117.67) -- (395,117.67) ;
%Straight Lines [id:da923818669339153] 
\draw [color={rgb, 255:red, 76; green, 120; blue, 54 }  ,draw opacity=1 ][fill={rgb, 255:red, 214; green, 242; blue, 190 }  ,fill opacity=1 ]   (364,157.67) -- (394,157.67) ;
%Shape: Rectangle [id:dp5357298361431284] 
\draw  [color={rgb, 255:red, 56; green, 160; blue, 216 }  ,draw opacity=1 ][fill={rgb, 255:red, 183; green, 230; blue, 255 }  ,fill opacity=1 ] (395,107.67) -- (455,107.67) -- (455,166.67) -- (395,166.67) -- cycle ;
%Straight Lines [id:da7535783823992097] 
\draw [color={rgb, 255:red, 76; green, 120; blue, 54 }  ,draw opacity=1 ][fill={rgb, 255:red, 214; green, 242; blue, 190 }  ,fill opacity=1 ]   (456,118.67) -- (487,118.67) ;
%Straight Lines [id:da3645269753665532] 
\draw  [color={rgb, 255:red, 76; green, 120; blue, 54 }  ,draw opacity=1 ][fill={rgb, 255:red, 214; green, 242; blue, 190 }  ,fill opacity=1 ]  (456,158.67) -- (486,158.67) ;
%Shape: Rectangle [id:dp28225474732210265] 
\draw  [color={rgb, 255:red, 148; green, 134; blue, 214 }  ,draw opacity=1 ][fill={rgb, 255:red, 216; green, 209; blue, 255 }  ,fill opacity=1 ] (487,108.67) -- (547,108.67) -- (547,167.67) -- (487,167.67) -- cycle ;
%Straight Lines [id:da5759008557039103] 
\draw  [color={rgb, 255:red, 76; green, 120; blue, 54 }  ,draw opacity=1 ][fill={rgb, 255:red, 214; green, 242; blue, 190 }  ,fill opacity=1 ]  (547,116.67) -- (578,116.67) ;
%Straight Lines [id:da46078845975732996] 
\draw [color={rgb, 255:red, 76; green, 120; blue, 54 }  ,draw opacity=1 ][fill={rgb, 255:red, 214; green, 242; blue, 190 }  ,fill opacity=1 ]   (547,156.67) -- (579,156.67) ;
%Shape: Rectangle [id:dp42646036953445754] 
\draw [color={rgb, 255:red, 76; green, 120; blue, 54 }  ,draw opacity=1 ][fill={rgb, 255:red, 214; green, 242; blue, 190 }  ,fill opacity=1 ]  (579,108.67) -- (639,108.67) -- (639,167.67) -- (579,167.67) -- cycle ;
%Curve Lines [id:da473428266576285] 
\draw    (591,144.67) .. controls (605,128.67) and (620,129.67) .. (631,144.67) ;
%Straight Lines [id:da28760086722089784] 
\draw    (622,127.67) -- (612,144.67) ;

% Text Node
\draw (72,128.4) node [anchor=north west][inner sep=0.75pt]    {$U_{\mathbf{2}^n}$};
% Text Node
\draw (150,128.4) node [anchor=north west][inner sep=0.75pt]    {$e^{-i\gamma_1H_C}$};
% Text Node
\draw (240,130.4) node [anchor=north west][inner sep=0.75pt]    {$e^{-i\beta_1H_B}$};
% Text Node
\draw (331,135) node [anchor=north west][inner sep=0.75pt]    {$\cdots $};
% Text Node
\draw (401,127.4) node [anchor=north west][inner sep=0.75pt]    {$e^{-i\gamma_pH_C}$};
% Text Node
\draw (491,129.4) node [anchor=north west][inner sep=0.75pt]    {$e^{-i\beta_pH_B}$};
% Text Node
\draw (40,122) node [anchor=north west][inner sep=0.75pt]    {$\vdots $};
% Text Node
\draw (121,122) node [anchor=north west][inner sep=0.75pt]    {$\vdots $};
% Text Node
\draw (211,122) node [anchor=north west][inner sep=0.75pt]    {$\vdots $};
% Text Node
\draw (302,122) node [anchor=north west][inner sep=0.75pt]    {$\vdots $};
% Text Node
\draw (371,122) node [anchor=north west][inner sep=0.75pt]    {$\vdots $};
% Text Node
\draw (462,122) node [anchor=north west][inner sep=0.75pt]    {$\vdots $};
% Text Node
\draw (555,122) node [anchor=north west][inner sep=0.75pt]    {$\vdots $};

\end{tikzpicture}
\label{approximate}
\caption{The quantum approximate optimization algorithm ansatz.  Preparation of the superposition of all $n$-bit strings is followed by $2p$ alternating quantum gates parameterized by variational classical parameters.  Then measurement of all qubits in the computational basis is performed, and the result is returned.}
\end{figure*}
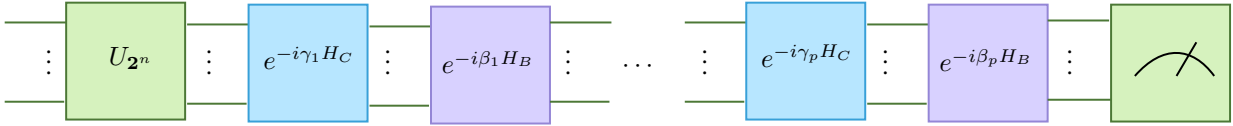

\section{Quantum Alternating Operator Ansatz}
\label{sec3}
Following the history of the field, we start by describing the quantum adiabatic algorithm (QAA) for optimization on quantum annealers.  This algorithm is discretized by the Trotter-Suzuki expansion~\cite{st_expansion} and approximated by unitary gates, yielding the quantum approximate optimization algorithm~\cite{qaoafarhi} (we call this Farhi's QAOA).  This algorithm is generalized for arbitrary problems, yielding the quantum alternating operator ansatz~\cite{hadfield2019} (just QAOA).  As mentioned in the introduction, the literature refers to both these algorithms as QAOA, as Farhi's QAOA algorithm is a special case of Hadfield's framework. We finish with a discussion of the classical and quantum parameter spaces for QAOA, and recent extensions.

\subsection{Quantum Adiabatic Algorithm}

Introduced by Farhi et al., the quantum adiabatic algorithm is designed to find the exact solution to an unconstrained optimization problem using a quantum annealer~\cite{qaa}.  In a separate paper~\cite{qaoafarhi}, Farhi et al. discuss the application of this approach to the maximum cut problem.  The idea is to encode the cost function $C$ into a matrix $H_C$ such that

\[H_C\ket{z}=C(z)\ket{z},\]

\noindent which is Hermitian since it is a diagonal real matrix.  A solution to the problem is clearly an eigenstate of $H_C$ with the largest eigenvalue.  Now initialize the system to 

\[\ket{z_0}=\frac{1}{2^n}\ket{+}^{\otimes n}=\frac{1}{2^n}\sum_\text{$n$-bit strings $z$}\ket{z},\]

\noindent which is a largest eigenstate of the transverse field Hamiltonian given by $H_B=\sum_{j=0}^{n-1} X_j$. Next, evolve the system from $H_0=H_B$ to $H_1=H_C$ so that by the adiabatic theorem it ends up in one of the highest energy eigenstates of $H_C$.  Measuring will collapse this state to a computational basis state corresponding to a bitstring maximizing the cost function. 
This argument requires that there always be a spectral gap in $H$ for all $t$, which is guaranteed by the Perron-Frobenius theorem for Metzler matrices~\cite{arrow1989}.  Now, we want a gate-based quantum algorithm approximating the QAA for a few reasons.  First, gate-based quantum computers are more general-purpose than quantum annealers.  Also, the QAA may take an arbitrary amount of time to run depending on the spectral gap between $H_B$ and $H_C$, which is undesirable.  For these reasons and other concerns about robustness, a gate-based analogue was developed, which we will discuss in the next section.

\subsection{Quantum Approximate Optimization Algorithm}\label{sec:qaoalgorithm}

In order to adapt the QAA to unitary gate based NISQ devices, Farhi et al. devised a discretization of the continuous adiabatic evolution and approximated it by Trotterization \cite{qaoafarhi}.  The resulting ansatz is called the quantum approximate optimization algorithm with $2p$ angles and has the form
\[e^{-i\gamma_p H_C}e^{-i\beta_p H_B}\cdots e^{-i\gamma_1 H_C}e^{-i\beta_1 H_B}\ket{z_0},\]

\noindent where $\gamma_1,\ldots,\gamma_p,\beta_1,\ldots,\beta_p$ are $2p$ real variational parameters called \textit{angles}.  For large enough $p$, with the correct parameters chosen, it is expected that a VQA with this ansatz will yield a solution. We will describe in detail how this algorithm is derived, and show how that theoretically guarantees correctness under certain assumptions about the parameter choices.  But we emphasize that an exact understanding of this derivation is not critical for designing implementations of this algorithm for particular problems or problem classes.

Let $Q(i,j)$ be the operator evolving a system by QAA from $t=i$ to $t=j$.  Then the a partition of the QAA process into $n$ time steps (which we will approximate on a gate based computer) is given by the composition of operators 
\[Q((n-1)/n,1)\cdots Q(0,1/n)\ket{z_0},\] 

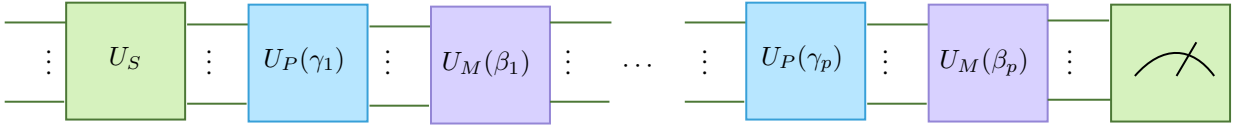
\begin{figure*}
    \centering
    \tikzset{every picture/.style={line width=0.75pt}} %set default line width to 0.75pt        

\begin{tikzpicture}[x=0.75pt,y=0.75pt,yscale=-1,xscale=1]

%Straight Lines [id:da22528067311212796] 
\draw  [color={rgb, 255:red, 76; green, 120; blue, 54 }  ,draw opacity=1 ][fill={rgb, 255:red, 214; green, 242; blue, 190 }  ,fill opacity=1 ]  (21,117.67) -- (52,117.67) ;
%Straight Lines [id:da9446647413192386] 
\draw  [color={rgb, 255:red, 76; green, 120; blue, 54 }  ,draw opacity=1 ][fill={rgb, 255:red, 214; green, 242; blue, 190 }  ,fill opacity=1 ]  (21,157.67) -- (51,157.67) ;
%Shape: Rectangle [id:dp2041100220624501] 
\draw  [color={rgb, 255:red, 76; green, 120; blue, 54 }  ,draw opacity=1 ][fill={rgb, 255:red, 214; green, 242; blue, 190 }  ,fill opacity=1 ]  (52,107.67) -- (112,107.67) -- (112,166.67) -- (52,166.67) -- cycle ;
%Straight Lines [id:da5584303350728332] 
\draw  [color={rgb, 255:red, 76; green, 120; blue, 54 }  ,draw opacity=1 ][fill={rgb, 255:red, 214; green, 242; blue, 190 }  ,fill opacity=1 ]  (113,118.67) -- (144,118.67) ;
%Straight Lines [id:da40668721267697694] 
\draw  [color={rgb, 255:red, 76; green, 120; blue, 54 }  ,draw opacity=1 ][fill={rgb, 255:red, 214; green, 242; blue, 190 }  ,fill opacity=1 ]  (113,158.67) -- (143,158.67) ;
%Shape: Rectangle [id:dp18065061971939755] 
\draw [color={rgb, 255:red, 56; green, 160; blue, 216 }  ,draw opacity=1 ][fill={rgb, 255:red, 183; green, 230; blue, 255 }  ,fill opacity=1 ]   (144,108.67) -- (204,108.67) -- (204,167.67) -- (144,167.67) -- cycle ;
%Straight Lines [id:da15392480213618565] 
\draw  [color={rgb, 255:red, 76; green, 120; blue, 54 }  ,draw opacity=1 ][fill={rgb, 255:red, 214; green, 242; blue, 190 }  ,fill opacity=1 ]  (205,119.67) -- (236,119.67) ;
%Straight Lines [id:da9262315254177915] 
\draw  [color={rgb, 255:red, 76; green, 120; blue, 54 }  ,draw opacity=1 ][fill={rgb, 255:red, 214; green, 242; blue, 190 }  ,fill opacity=1 ]  (205,159.67) -- (235,159.67) ;
%Shape: Rectangle [id:dp0928579716614486] 
\draw  [color={rgb, 255:red, 148; green, 134; blue, 214 }  ,draw opacity=1 ][fill={rgb, 255:red, 216; green, 209; blue, 255 }  ,fill opacity=1 ] (236,109.67) -- (296,109.67) -- (296,168.67) -- (236,168.67) -- cycle ;
%Straight Lines [id:da048330393188396714] 
\draw [color={rgb, 255:red, 76; green, 120; blue, 54 }  ,draw opacity=1 ][fill={rgb, 255:red, 214; green, 242; blue, 190 }  ,fill opacity=1 ]   (296,117.67) -- (327,117.67) ;
%Straight Lines [id:da21812976650155824] 
\draw [color={rgb, 255:red, 76; green, 120; blue, 54 }  ,draw opacity=1 ][fill={rgb, 255:red, 214; green, 242; blue, 190 }  ,fill opacity=1 ]   (296,157.67) -- (326,157.67) ;
%Straight Lines [id:da6949188181915573] 
\draw  [color={rgb, 255:red, 76; green, 120; blue, 54 }  ,draw opacity=1 ][fill={rgb, 255:red, 214; green, 242; blue, 190 }  ,fill opacity=1 ]  (364,117.67) -- (395,117.67) ;
%Straight Lines [id:da923818669339153] 
\draw  [color={rgb, 255:red, 76; green, 120; blue, 54 }  ,draw opacity=1 ][fill={rgb, 255:red, 214; green, 242; blue, 190 }  ,fill opacity=1 ]  (364,157.67) -- (394,157.67) ;
%Shape: Rectangle [id:dp5357298361431284] 
\draw  [color={rgb, 255:red, 56; green, 160; blue, 216 }  ,draw opacity=1 ][fill={rgb, 255:red, 183; green, 230; blue, 255 }  ,fill opacity=1 ] (395,107.67) -- (455,107.67) -- (455,166.67) -- (395,166.67) -- cycle ;
%Straight Lines [id:da7535783823992097] 
\draw  [color={rgb, 255:red, 76; green, 120; blue, 54 }  ,draw opacity=1 ][fill={rgb, 255:red, 214; green, 242; blue, 190 }  ,fill opacity=1 ]  (456,118.67) -- (487,118.67) ;
%Straight Lines [id:da3645269753665532] 
\draw  [color={rgb, 255:red, 76; green, 120; blue, 54 }  ,draw opacity=1 ][fill={rgb, 255:red, 214; green, 242; blue, 190 }  ,fill opacity=1 ]  (456,158.67) -- (486,158.67) ;
%Shape: Rectangle [id:dp28225474732210265] 
\draw  [color={rgb, 255:red, 148; green, 134; blue, 214 }  ,draw opacity=1 ][fill={rgb, 255:red, 216; green, 209; blue, 255 }  ,fill opacity=1 ] (487,108.67) -- (547,108.67) -- (547,167.67) -- (487,167.67) -- cycle ;
%Straight Lines [id:da5759008557039103] 
\draw  [color={rgb, 255:red, 76; green, 120; blue, 54 }  ,draw opacity=1 ][fill={rgb, 255:red, 214; green, 242; blue, 190 }  ,fill opacity=1 ]  (547,116.67) -- (578,116.67) ;
%Straight Lines [id:da46078845975732996] 
\draw  [color={rgb, 255:red, 76; green, 120; blue, 54 }  ,draw opacity=1 ][fill={rgb, 255:red, 214; green, 242; blue, 190 }  ,fill opacity=1 ]  (547,156.67) -- (579,156.67) ;
%Shape: Rectangle [id:dp42646036953445754] 
\draw [color={rgb, 255:red, 76; green, 120; blue, 54 }  ,draw opacity=1 ][fill={rgb, 255:red, 214; green, 242; blue, 190 }  ,fill opacity=1 ] (579,108.67) -- (639,108.67) -- (639,167.67) -- (579,167.67) -- cycle ;
%Curve Lines [id:da473428266576285] 
\draw    (591,144.67) .. controls (605,128.67) and (620,129.67) .. (631,144.67) ;
%Straight Lines [id:da28760086722089784] 
\draw    (622,127.67) -- (612,144.67) ;

% Text Node
\draw (72,128.4) node [anchor=north west][inner sep=0.75pt]    {$U_{S}$};
% Text Node
\draw (150,128.4) node [anchor=north west][inner sep=0.75pt]    {$U_{P}( \gamma _{1})$};
% Text Node
\draw (240,130.4) node [anchor=north west][inner sep=0.75pt]    {$U_{M}( \beta _{1})$};
% Text Node
\draw (331,135) node [anchor=north west][inner sep=0.75pt]    {$\cdots $};
% Text Node
\draw (401,127.4) node [anchor=north west][inner sep=0.75pt]    {$U_{P}( \gamma _{p})$};
% Text Node
\draw (491,129.4) node [anchor=north west][inner sep=0.75pt]    {$U_{M}( \beta _{p})$};
% Text Node
\draw (40,122) node [anchor=north west][inner sep=0.75pt]    {$\vdots $};
% Text Node
\draw (121,122) node [anchor=north west][inner sep=0.75pt]    {$\vdots $};
% Text Node
\draw (211,122) node [anchor=north west][inner sep=0.75pt]    {$\vdots $};
% Text Node
\draw (302,122) node [anchor=north west][inner sep=0.75pt]    {$\vdots $};
% Text Node
\draw (371,122) node [anchor=north west][inner sep=0.75pt]    {$\vdots $};
% Text Node
\draw (462,122) node [anchor=north west][inner sep=0.75pt]    {$\vdots $};
% Text Node
\draw (555,122) node [anchor=north west][inner sep=0.75pt]    {$\vdots $};

\end{tikzpicture}
\label{alternating}
\caption{The QAOA$_p$ ansatz.  Analogous to the quantum approximate optimization algorithm, initial state preparation is followed by $2p$ alternating quantum gates parameterized by variational classical parameters.  Then measurement of all qubits in the computational basis is performed, and the result is returned.}
\label{fig:my_label}
\end{figure*}

\noindent although in general it is hard to calculate the evolution of quantum systems.  But when the Hamiltonian $H$ is time-independent, we have simply the time-independent evolution operator $U(H,t)=\exp(-iHt)$ (with units normalized by the reduced Planck constant).  For the quantum approximate optimization algorithm, the idea is to take large enough $p$ so
\begin{align}
    Q(i/p,(i+1)/p)&\approx U(H(i/p),1/p)\\ \nonumber
    &=\exp(-iH(i/p)/p).
\end{align}

\noindent for $H$ a time-dependent Hamiltonian adiabatically evolving the system from $H_B$ to $H_C$.  Applied slowly enough, the \textit{linear annealing schedule} $H(t)=tH_B+(1-t)H_C$ performs such an evolution. Then using equation 1 repeatedly, we can compute
\[Q((p-1)/p,1/p)Q((p-2)/p,(p-1)/p)\cdots Q(0,1/p)\ket{z_0}\]
\[\approx U(H((p-1)/p),1/p)\cdots U(H(0),1/p)\ket{z_0}\]
\[=\exp(-iH((p-1)/p)/p)\cdots \exp(-iH(0)/p)\ket{z_0}.\]

Now recall the \textit{Trotter-Suzuki expansion} (sometimes called \textit{Trotterization}, or the \textit{Lie product formula}) \begin{equation}
    e^{x(A+B)}=e^Ae^B+\mathcal{O}(x^2)
\end{equation}
\noindent for non-commuting operators $A,B$~\cite{st_expansion}.  Using this equation~2, we can approximate our expression by computing
\begin{align*}
    & \prod_{j=0}^{p-1}\exp(-iH(j/p)/p) \\
    &=\prod_{j=0}^{p-1}\exp\left(\frac{-i}{p}\Big(jH_C/p+(1-j/p)H_B\Big)\right) \\
    &=\prod_{j=0}^{p-1}\Big(\exp(-ijH_C/p^2)\exp(-i(1-j/p)H_B/p)+\mathcal{O}(1/p^2)\Big) \\
    &\approx\prod_{j=1}^{p}\Big(\exp(-i\gamma_jH_C)\exp(-i\beta_jH_B)\Big),
\end{align*}
\noindent with convergence in the limit as $p$ goes to infinity, for small values $\beta_j,\gamma_j$, and where we use the notation
\[\prod_{i=0}^{n} A_i=A_n\circ A_{n-1}\circ\cdots\circ A_1\circ A_0\] for simplicity.  This computation results in the ansatz for Farhi's QAOA, given $H_B$ and $H_C$ as in the previous section.  We should note here that $\exp(-i\gamma_j H_C)$ and $\exp(-i\beta_j H_B)$ are unitary because $-i\gamma_j H_C$ and $-i\beta_j H_B$ are Hermitian.  See Fig.~2 for a quantum circuit diagram for Farhi's QAOA.

Based on its design, this algorithm should theoretically converge to the maximum in the limit of large $p$.  That is, the quantum expectation value 
$\braket{QAOA|H_C|QAOA} \stackrel{p\to \infty}{\to}1$.  
However, for a small number of rounds, such as normally considered, optimal witness angles with proven approximation guarantees~\cite{wurtz2021,caha2022} do not have to come from a discretization \& Trotterization of the adiabatic algorithm.
Furthermore, the above guarantee is conditional on some notable assumptions:  First, it assumes the ability to pick optimal angles and arbitrarily many reliable qubits. Also, it assumes that the QAA would solve this case; for example, this requires there is always a spectral gap in $H_t$.  Finally we used Trotterization, so we assumed that $A$ and $B$ do not commute. These assumptions are important to consider in practice.  Furthermore, this algorithm is designed specifically for unconstrained optimization problems.   In the next section, we discuss a metaheuristic called the quantum alternating operator ansatz which addresses these and other shortcomings of this quantum approximate optimization algorithm, making it more generally applicable.

\subsection{Quantum Alternating Operator Ansatz}

In order to extend the quantum approximate optimization algorithm to constrained optimization problems and produce an overall more robust ansatz, Hadfield et al. introduced the quantum alternating operator ansatz (QAOA)~\cite{hadfield2019}.  This ansatz generalizes on the form of Farhi's QAOA, focusing on three  \textit{quantum parameters}, which we will briefly describe:

\begin{enumerate}
    \item The \textit{phase separator operators} $U_P(\gamma)$, generalize the problem separators $e^{-i\gamma H_C}$.  These should be diagonal in the computational basis, and efficient to produce.  Also, good solutions should have relatively large eigenvalues separated by a spectral gap from bad solutions. 
    \item The \textit{mixing operators} $U_M(\beta)$ generalize the transverse field mixer $e^{-i\beta H_B}$.  The search space $S$ must be invariant under this family, and all feasible states $x,y\in S$ must satisfy the \textit{transition condition}
    \begin{align}|\bra{x}U_M^r(\beta)\ket{y}|>0\end{align}
    
    \noindent for some positive integer $r$ and parameter value $\beta$.  Interpreting the Dirac bra-ket notation, this expression means that there is a nonzero probability that measuring $U_M^r(\beta)\ket{t}$ yields $\ket{s}$. These conditions explain the intuitive notion this operator ``mixes'' the search space.  We ideally want to be able to efficiently build this gate, and do not want it to commute with $U_P(\gamma)$ in general.
    \item The \textit{initial state operator} $U_S$ produces the initial state $\ket{z_0}$.  This state is usually chosen to represent the search space or some subset of it.  For example, in Farhi's QAOA, $U_S$ produces the superposition of $S=\mathbf{2}^n$, the entire unconstrained search space. Any such operator $U_S$ should  be \textit{trivial} in that it should be a small constant depth circuit w.r.t. $n$.  In particular, Hadfield et al. caution that relaxations of this triviality condition much beyond logarithmic ``would obviate the usefulness of the ansatz ... and should be considered hybrid algorithms, with an initialization part and a QAOA part''~\cite{hadfield2019}. However, such hybrid algorithms may still be of independent interest.
\end{enumerate}

\begin{figure*}
\begin{center}
\begin{tikzpicture}[x=0.75pt,y=0.75pt,yscale=-1,xscale=1]

%Flowchart: Process [id:dp014803447420872695] 
\draw  [color={rgb, 255:red, 172; green, 57; blue, 49 }  ,draw opacity=1 ][fill={rgb, 255:red, 255; green, 166; blue, 158 }  ,fill opacity=1 ] (255,30) -- (490.78,30) -- (490.78,271) -- (255,271) -- cycle ;
%Flowchart: Process [id:dp4545900342986313] 
\draw  [color={rgb, 255:red, 76; green, 120; blue, 54 } ,draw opacity=1 ][fill={rgb, 255:red, 214; green, 242; blue, 190 }  ,fill opacity=1 ] (343,56) -- (399.78,56) -- (399.78,105.68) -- (343,105.68) -- cycle ;
%Flowchart: Process [id:dp04469188432352067] 
\draw  [color={rgb, 255:red, 244; green, 145; blue, 0 }   ,draw opacity=1 ][fill={rgb, 255:red, 253; green, 196; blue, 95 }   ,fill opacity=1 ] (267,122) -- (478.78,122) -- (478.78,237) -- (267,237) -- cycle ;
%Flowchart: Process [id:dp03232301999702214] 
\draw  [color={rgb, 255:red, 56; green, 160; blue, 215 }  ,draw opacity=1 ][fill={rgb, 255:red, 183; green, 230; blue, 255 }  ,fill opacity=1 ] (397,156) -- (453.78,156) -- (453.78,205.68) -- (397,205.68) -- cycle ;
%Flowchart: Process [id:dp7326300292091998] 
\draw  [color={rgb, 255:red, 148; green, 134; blue, 214 }  ,draw opacity=1 ][fill={rgb, 255:red, 216; green, 209; blue, 255 }  ,fill opacity=1 ] (292,155) -- (348.78,155) -- (348.78,204.68) -- (292,204.68) -- cycle ;

%Bend Arrow [id:dp9732431412686593] 
\draw  [color={rgb, 255:red, 76; green, 120; blue, 54 }  ,draw opacity=1 ][fill={rgb, 255:red, 214; green, 242; blue, 190 }  ,fill opacity=1 ] (408.14,76.14) -- (414.07,76.14) .. controls (418.97,76.14) and (422.95,80.12) .. (422.95,85.03) -- (422.94,138) -- (424.86,138) -- (420.43,147.14) -- (416,138) -- (417.92,138) -- (417.93,85.03) .. controls (417.93,82.89) and (416.2,81.16) .. (414.07,81.16) -- (408.14,81.16) -- cycle ;
%Flowchart: Alternative Process [id:dp9792020547424378] 
\draw  [color={rgb, 255:red, 76; green, 120; blue, 54 }  ,draw opacity=1 ][fill={rgb, 255:red, 214; green, 242; blue, 190 }  ,fill opacity=1 ] (153.78,137.28) .. controls (153.78,132.15) and (157.93,128) .. (163.05,128) -- (226.73,128) .. controls (231.85,128) and (236,132.15) .. (236,137.28) -- (236,171.73) .. controls (236,176.85) and (231.85,181) .. (226.73,181) -- (163.05,181) .. controls (157.93,181) and (153.78,176.85) .. (153.78,171.73) -- cycle ;
%Bend Arrow [id:dp9732431412686593] 
\draw  [color={rgb, 255:red, 76; green, 120; blue, 54 }  ,draw opacity=1 ][fill={rgb, 255:red, 214; green, 242; blue, 190 }  ,fill opacity=1 ] (408.14,76.14) -- (414.07,76.14) .. controls (418.97,76.14) and (422.95,80.12) .. (422.95,85.03) -- (422.94,138) -- (424.86,138) -- (420.43,147.14) -- (416,138) -- (417.92,138) -- (417.93,85.03) .. controls (417.93,82.89) and (416.2,81.16) .. (414.07,81.16) -- (408.14,81.16) -- cycle ;
%Flowchart: Alternative Process [id:dp9792020547424378] 
\draw  [color={rgb, 255:red, 76; green, 120; blue, 54 }  ,draw opacity=1 ][fill={rgb, 255:red, 214; green, 242; blue, 190 }  ,fill opacity=1 ] (153.78,137.28) .. controls (153.78,132.15) and (157.93,128) .. (163.05,128) -- (226.73,128) .. controls (231.85,128) and (236,132.15) .. (236,137.28) -- (236,171.73) .. controls (236,176.85) and (231.85,181) .. (226.73,181) -- (163.05,181) .. controls (157.93,181) and (153.78,176.85) .. (153.78,171.73) -- cycle ;
%Bend Arrow [id:dp05965673692204221] 
\draw  [color={rgb, 255:red, 76; green, 120; blue, 54 }  ,draw opacity=1 ][fill={rgb, 255:red, 214; green, 242; blue, 190 }  ,fill opacity=1 ] (189.78,120) -- (189.78,102.14) .. controls (189.78,87.35) and (201.77,75.36) .. (216.56,75.36) -- (324,75.36) -- (324,73) -- (334.78,77.75) -- (324,82.5) -- (324,80.14) -- (216.56,80.14) .. controls (204.41,80.14) and (194.56,89.99) .. (194.56,102.14) -- (194.56,120) -- cycle ;
%Bend Arrow [id:dp6891971860985986] 
\draw  [color={rgb, 255:red, 76; green, 120; blue, 54 }  ,draw opacity=1 ][fill={rgb, 255:red, 214; green, 242; blue, 190 }  ,fill opacity=1 ] (297.72,220.1) -- (205.21,219.92) .. controls (197.45,219.91) and (191.18,213.61) .. (191.2,205.86) -- (191.22,196.5) -- (188.77,196.5) -- (193.79,186.92) -- (198.76,196.52) -- (196.32,196.51) -- (196.3,205.87) .. controls (196.29,210.8) and (200.28,214.81) .. (205.22,214.82) -- (297.73,215) -- cycle ;
%Left Right Arrow [id:dp13177913768199123] 
\draw  [color={rgb, 255:red, 76; green, 120; blue, 54 }  ,draw opacity=1 ][fill={rgb, 255:red, 214; green, 242; blue, 190 }  ,fill opacity=1 ] (351.89,180) -- (361,176) -- (361,177.9) -- (384.78,177.9) -- (384.78,176) -- (393.89,180) -- (384.78,184) -- (384.78,182.1) -- (361,182.1) -- (361,184) -- cycle ;

\draw (260,35) node [anchor=north west][inner sep=0.75pt]    {\texttt{Algorithm 1}};

\draw (150,35) node [anchor=north west][inner sep=0.75pt]    {\texttt{Algorithm 2}};

% Text Node
\draw (360,72.4) node [anchor=north west][inner sep=0.75pt]    {$U_{S}$};
% Text Node
\draw (401,172.4) node [anchor=north west][inner sep=0.75pt]    {$U_{P}( \gamma _{i})$};
% Text Node
\draw (295,171.4) node [anchor=north west][inner sep=0.75pt]    {$U_{M}( \beta _{i})$};
% Text Node
\draw (162,140) node [anchor=north west][inner sep=0.75pt]   [align=left] {{\fontfamily{pcr}\selectfont Optimize}};
% Text Node
\draw (395,219) node [anchor=north west][inner sep=0.75pt]  [font=\scriptsize] [align=left] {$\displaystyle p$};
% Text Node
\draw (405,216) node [anchor=north west][inner sep=0.75pt]  [font=\footnotesize] [align=left] {{\fontfamily{pcr}\selectfont iterations}};
% Text Node
\draw (400,250) node [anchor=north west][inner sep=0.75pt]  [font=\footnotesize] [align=left] {{\fontfamily{pcr}\selectfont QAOA$_p$ ansatz}};
% Text Node
\draw (196.89,157.9) node [anchor=north west][inner sep=0.75pt]  [font=\small]  {$\boldsymbol\beta ,\boldsymbol\gamma $};
% Text Node
\draw (165,157.9) node [anchor=north west][inner sep=0.75pt]  [font=\small]  {$C$};

\draw (178,157.9) node [anchor=north west][inner sep=0.75pt]    {\fontfamily{pcr}\selectfont by };
\end{tikzpicture}
\caption{A process diagram for the entire QAOA VQA (Algorithm 2).  The red outer box represents the QAOA ansatz (Algorithm 1) as a whole while the inner orange box represents the  alternating operator approximation (Trotterization) of the QAA adiabatic process.  This ansatz is fed into an overarching classical optimization loop, which runs until some convergence criterion is met, or for a specified maximum number of iterations in the case that it fails to halt.}
\end{center}
\end{figure*}
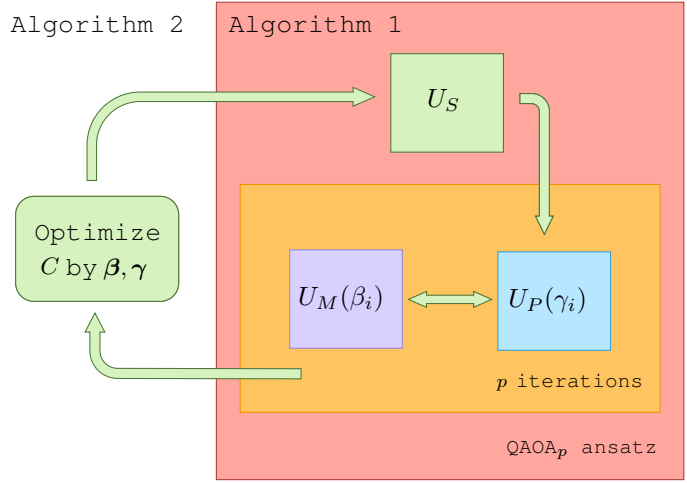

It is clear that the quantum approximate optimization algorithm satisfies these conditions.  First, the phase separator $U_P(\gamma_j)=\exp(-i\gamma_j H_C)$ is diagonal since $H_C$ is diagonal.  Next, the mixer $U_M(\beta_j)=\exp(-i\beta_j H_B)$ sends $\ket{x}$ to $\ket{y}$ with some probability (for $\beta_j\neq 0$) since $H_B$ does with probability $1/2^n$.  Thus we can take $r=1$ for the transition condition, equation 3 above.  Finally, $U_S=H^{\otimes n}$ produces the state $\ket{z_0}=\ket{+}^{\otimes n}$ in $\mathcal{O}(1)$ with respect to $n$.

It's desirable for operators to be exponents of the form $\exp(-i\alpha \sum_j a_jH_j)$, for commutative $H_j$, allowing for implementation using unitary gates with the form $\exp(-i\alpha a_jH_j)$.   

In any event, these parameters form the QAOA$_p$ ansatz
\[U_M(\beta_p)U_P(\gamma_p)\cdots U_M(\beta_1)U_P(\gamma_1)U_S\]

\noindent by analogy with the quantum approximate optimization algorithm.  See Fig. 3 for a schematic diagram of this quantum circuit, and Algorithm 1 for a pseudocode implementation.

 Other critical design choices go into this algorithm, which we call \textit{classical parameters} and summarize below:
\begin{itemize}
    \item The \textit{number of angles} $2p$.  This somewhat innocuous choice conceals a design paradox:  the algorithm should theoretically run better with larger $p$; however larger $p$ increases the circuit depth, exposing the computation to a higher risk for NISQ device quantum errors.
    \item The \textit{optimization algorithm} used for the classical part of the VQA can critically affect the performance.  This includes a \textit{convergence criterion} to stop the algorithm after a certain point.  In the case a theoretical optimum $x$ is not known, you can wait for the result to converge to some small error. Also, a maximum number of iteration \textit{max\_iter} should be included to ensure the program halts.
    \item The \textit{problem encoding} affects the choices for all other parameters.  For example, toy optimization problems with just two feasible solutions can be encoded as one bit with the \textit{binary encoding} $\{0,1\}$ or as two bits with the \textit{one-hot encoding} $\{10,01\}$.  For more discussion see \cite{fuchs2021}.
\end{itemize}

\begin{algorithm}[b]
\caption{The QAOA quantum ansatz}

\ \newline\textbf{input: } Operators $U_S$, $U_M$, $U_P$.

\textbf{output: } A bitstring $z$.\newline

Prepare the initial state $\ket{z_0}$ using the operator $U_S$;
%\footnote{Here we digress to follow the formalism of B\"artschi \textit{et al}~\cite{baertschi2020grover}, which explicitly considers the state preparation as part of the ansatz.}

\textbf{for} each number $i$ with $1\leq i\leq p$, in order, \textbf{do}

\hspace{1cm} Apply the operator $U_M(\beta_i)\circ U_P(\gamma_i)$; 

\textbf{end for}

\textit{measure} = Measure the qubits;

\textbf{return} \textit{measure};\newline
\end{algorithm}

Choosing an optimization algorithm allows us to construct the variational part of the algorithm, given in Algorithm 2.  Taking all these choices together gives us the entire QAOA VQA, depicted by a process diagram in Fig. 4.

We organize these quantum and classical parameters for QAOA in Fig's. 5 and 6 respectively, with references to papers working with different options. Designing a QAOA instance for a problem involves selecting values for the parameters given in these figures. Algorithm 2 and Fig. 4 detail how these parts fit together into a cohesive unit.  Algorithm 2 details the VQA structure wrapping Algorithm 1, and Fig. 4 shows the whole process in a diagram distinguishing the two algorithms. Many choices are possible, and many have been studied: Many choices are possible, and many have been studied: Hadfield et al. provide a ``compendium of mappings and compilation'' as a guide for many problems~\cite{hadfield2019}.  This broad applicability across different classes of problems (e.g. constrained, unconstrained, permutation-based, etc.) is why we call QAOA a metaheuristic.

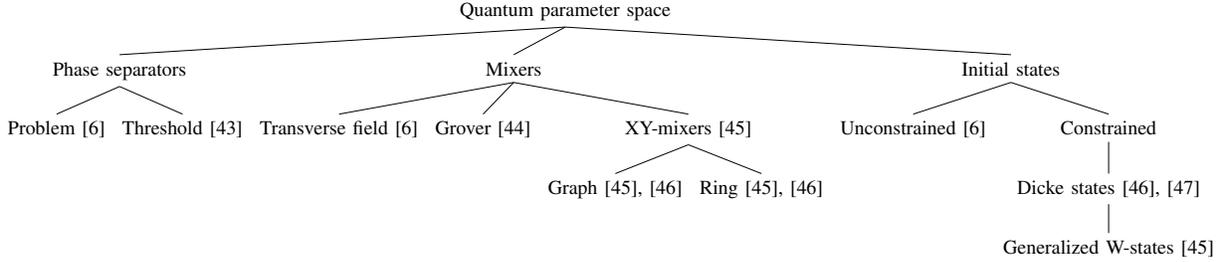
\begin{figure*}
\centering
\begin{tikzpicture}[scale=0.75]
\Tree[.{Quantum parameter space} 
    [.{Phase separators}
        [.{Problem~\cite{qaoafarhi} } ]
        [.{Threshold~\cite{golden2021threshold} } ] ]
    [.{Mixers} 
        [.{Transverse field~\cite{qaoafarhi}} ]
        [.{Grover~\cite{baertschi2020grover}} ]
        [.{XY-mixers~\cite{wang2020}}
            [.{Graph~\cite{wang2020,cook2020vertexcover}} ]
            [.{Ring~\cite{wang2020,cook2020vertexcover}} ] ] ] 
    [.{Initial states} 
        [.{Unconstrained~\cite{qaoafarhi} } ]
        [.{Constrained}
        [.{Dicke states~\cite{baertschi2019deterministic,cook2020vertexcover}} 
        [.{Generalized W-states~\cite{wang2020}} ] ] ] ] ]
\end{tikzpicture}
\caption{An overview of the quantum parameter space with references to works studying various specific choices for the parameters.  Threshold phase separators are designed to find a solution above a certain value; together with the Grover mixer this variant generalizes Grover's algorithm.  XY-mixers preserve the Hamming weights of superposed states.  The superposition of all Hamming weight $k$ states is called a Dicke state, and a generalized $W$-state when $k=1$.}
\end{figure*}

\begin{algorithm}[t]
\caption{The QAOA variational part}

\ \newline \textbf{input: } Operators $U_S$, $U_M$, $U_P$, initial angles $\boldsymbol\beta_0,\boldsymbol\gamma_0$, an optimization algorithm, a maximum number of iterations, and a function $C$ implementing the cost function f for the problem.

\textbf{output: }A distribution of feasible solutions.\newline

\textbf{let} $\boldsymbol\beta=\boldsymbol\beta_0$, $\boldsymbol\gamma=\boldsymbol\gamma_0$, $\textit{convergence}=0$, $\textit{num\_iter}=0$;

\textbf{while} no \textit{convergence} \textbf{and} \textit{num\_iter} $<$ \textit{max\_iter}, \textbf{do}

\hspace{1cm} \textbf{for} each number $i$ with $1\leq i\leq\textit{num\_trials}$, \textbf{do}

\hspace{2cm} \textit{measurement} = perform Algorithm 1;

\hspace{2cm} \textit{measurements}.append(\textit{measurement});

\hspace{1cm} \textbf{end for}
 
\hspace{1cm} check\_convergence(f, \textit{measurements});

\hspace{1cm} \textbf{if} no \textit{convergence}, \textbf{then}

\hspace{2cm} Optimize $\boldsymbol\beta,\boldsymbol\gamma$ w.r.t. $C$

\hspace{2cm} $\textit{num\_iter}=\textit{num\_iter}+1$;

\hspace{1cm} \textbf{end if}

\textbf{end while}

\textbf{return} \textit{measurements};\newline

\end{algorithm}

\subsection{Recent Directions and Extensions}

Since Hadfield et al.~\cite{hadfield2019}, several further generalizations and modifications of the QAOA metaheuristic have been proposed.  In this section we will summarize some of these, as well as discuss some recent directions in research involving QAOA.

\subsubsection{Some modified QAOA ans\"atze} The QAOA ansatz parameterizes each quantum gate with one classical angle.  Herrman et al. propose a multi-angle (ma-QAOA) ansatz in which operators $U(\beta)=\exp(-i\beta \sum_j U_j)$ are parameterized with many angles as $\exp(-i\sum_j\beta_j U_j)$ ~\cite{Herrman2022}.  Their empirical study showed many of these angles become zero,  simplifying the depth of the ansatz circuit.  This should help reduce NISQ errors, at the cost of a more involved optimization step.

Inspired by the adaptive VQE of Grimsley et al.~\cite{Grimsley2019}, Zhu et al. introduced an adaptive QAOA~\cite{zhu2022}.  This algorithm iteratively and automatically produces a QAOA ansatz tailored to a given problem.  They showed that their algorithm is better in various ways than the standard QAOA on a class of MaxCut instances, and explain this result with ``shortcuts to adiabaticity''.  Improvements include a reduced number of variational parameters and CNOT gates.

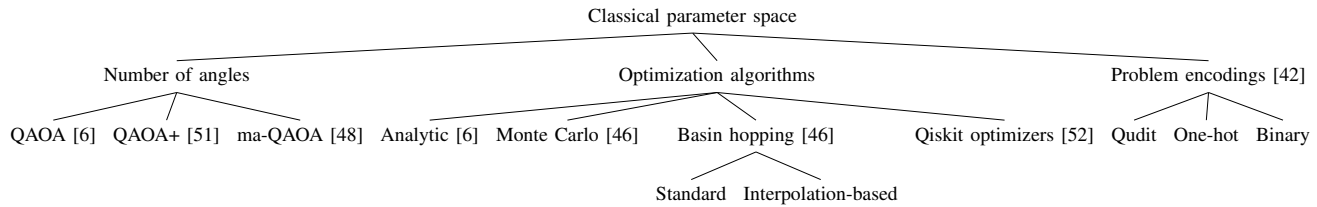
\begin{figure*}[t]
\begin{tikzpicture}[scale=0.75]
\Tree[.{Classical parameter space} 
    [.{Number of angles}  
        [.QAOA~\cite{qaoafarhi} ]
        [.QAOA+~\cite{Chalupnik2022} ]
        [.ma-QAOA~\cite{Herrman2022} ] ]
    [.{Optimization algorithms}
        [.{Analytic~\cite{qaoafarhi}} ]
        [.{Monte Carlo~\cite{cook2020vertexcover} } ] 
        [.{Basin hopping~\cite{cook2020vertexcover}} 
            [.{Standard} ]
            [.{Interpolation-based} ] ] 
        [.{Qiskit optimizers~\cite{fp2022}} ] ]
    [.{Problem encodings~\cite{fuchs2021}} 
        [.{Qudit} ]
        [.{One-hot} ]
        [.{Binary} ] ] ]
\end{tikzpicture}
\caption{An overview of the classical parameter space with references to works studying various specific choices for the parameters.  The usual QAOA ansatz has a parameter $p$ for the number of angles while QAOA+ and ma-QAOA add more variational angles.  The study of optimization algorithm choice is mostly numerical, while problem encodings are more important to the expression of an analytic formulation for a particular combinatorial optimization problem.}
\end{figure*}

An approach for choosing the initial state is to \textit{warm-start} QAOA by solving an easier, relaxed form of the problem and using such a solution as the initial state.  Introducing this idea, Egger et al~\cite{egger} relaxed the NP-hard quadratic unconstrained binary optimization (QUBO) problem to a convex quadratic program, or sometimes even semidefinite program to obtain an initial state.  This state was then used in the QAOA instance for some QUBO problems such as portfolio optimization (PO) and MaxCut, demonstrating better outcomes in some cases.

\subsubsection{Some recent QAOA research}  One focus of current research is the performance of certain parameters on various problems.  For example, B\"artschi et al. introduced the \textit{Grover mixer}, which is produced in a natural way using the initial state operator $U_S$ for the superposition of all feasible states~\cite{baertschi2020grover}.  In particular, they defined their mixer by $U_M(\beta)=\exp(-i\beta \ket{z_0}\bra{z_0})$ and showed it to work well for certain problems, offloading mixer design complexity to initial state preparation.  B\"artschi et al. also worked on constrained optimization, where they study the superposition of Hamming weight one states called \textit{Dicke states}~\cite{cook2020vertexcover,baertschi2019deterministic}.  They gave a linear depth circuit for Dicke state preparation and applied it to the max $k$-vertex cover problem. Their approach may work well for other Hamming weight constrained problems, but that requires more study to qualify.  Combining these results yields a mixer for Hamming weight constrained problems.

Other work focuses on rigorously proving facts about QAOA.  For example, Farhi and Harrow demonstrated how sampling QAOA circuits can demonstrate quantum supremacy~\cite{qaoasupremacy}.  More broadly, Hadfield et al. developed an analytic framework for analyzing QAOA and used it to demonstrate that QAOA always beats random guessing~\cite{qaoaanalytic}.

Of course there are many more modifications and applications than we can reasonably discuss in this article, and more appear every week.  We hope that these examples reasonably exemplify the state of the art in QAOA research, and provide helpful entry points for researchers interested in the field as a whole.  In the next section we will focus more specifically on topics relevant to computational intelligence.

\section{Connections to Computational Intelligence}

In this section, we first discuss how well-known computational intelligence techniques such as ant colony optimization and genetic algorithms can be used to improve QAOA. In the same vein, we then also discuss how QAOA, viewed as a computational intelligence technique, can substitute and complement these well-known metaheuristics. 

Ant colony optimization is used to simplify problems for QAOA and to reliably compile quantum circuits.  Genetic algorithms do a better job of this compilation, and reinforcement learning can be used to train transferable QAOA policies. Also, due to their robustness compared to classical gradient descent, these and other computational intelligence techniques can be used to avoid barren plateaus--a pathology characteristic to VQAs where vanishing gradients traps the optimizer in a suboptimal region of the search space~\cite{Wang2021}. This is a topic meriting further research attention due to the importance of variational quantum algorithms.

The main advantage of QAOA over these classical methods is the potential for guaranteed convergence to theoretically optimal solutions, although that behaviour corresponds to the limit as the number of angles goes to infinity.  Therefore this guarantee may not be feasibly exploitable on NISQ devices, and a more careful comparison of optimization methods is warranted for each particular problem.   Even when the QAOA may not  converge, it can still serve as a useful heuristic for finding near-optimal solutions.  

\label{sec4}

\subsection{Ant Colony Optimization}
\label{sec:aco}
Introduced by Dorigo et al., ant colony optimization (ACO) is a metaheuristic based on the swarm intelligence of ants~\cite{ants_first}.  Its most intuitive  application is to the traveling salesperson problem and its generalizations, with simulated ants `exploring' the problem graph.  Some applications and history of the algorithm is given in the survey by Dorigo et al.~\cite{antscim}

Some recent developments in ACO have centered on applications to multi-objective and continuous optimization~\cite{acoauto, acocont}, and improvements via parallelization~\cite{acoparallel}.  In parallel multi-colony optimization, threads run their own ant colonies and exchange pheromone information based on an overarching inter-colony \textit{exchange policy}.  This technique appears in emerging work on the application of ACO to QAOA.

Ghimire et al. approach the traveling salesperson problem (TSP) on NISQ devices with QAOA, using parallel ACO to break the graph into small enough pieces for noisy qubits to work on~\cite{acoqaoa}.   They find encouraging results, usually staying within 1\% of benchmark values for medium size graphs ($n\leq 150$) and 16 threads.  This hybrid technique should be studied more generally, as it may to other classes of problems.  

Furthermore, parallelization of quantum TSP is not the only possible application of ACO to QAOA. Baioletti et al. introduce an ACO algorithm (QCC-ACO) for efficient and scalable compilation of quantum circuits~\cite{qccaco}.  This approach should be tested on QAOA circuits due to the close association between these algorithms, and QAOA's role in near-term quantum computing as a recognized benchmark~\cite{benchmark}.

\subsection{Reinforcement Learning}

Policy gradient methods (PGMs) for reinforcement learning (RL) are current approaches for intelligent agents to find optimal policies for decision-making. These approaches are effective in combination with QAOA since they help avoid getting trapped in local minima by reserving some probability of jumping out of such traps.  The OpenAI proximal policy optimization (PPO) algorithm, introduced by Schulman et al., is a strong contender within this space which perform ``multiple epochs of minibatch updates'' per sample~\cite{reinforcementqaoa}.  Benchmarked against existing PGMs with single updates, PPO generally outperforms and runs more smoothly.

Khairy et al.~\cite{reinforcementqaoa} and Wauters et al.~\cite{qaoareinforcement} apply PPO to QAOA.  Khairy et al. study the Farhi's QAOA on MaxCut, and show their approach allows for training on small instances that is often transferable to larger cases.  More broadly, Wauters et al. show that a PPO approach learns the optimal policy for finding the ground state of the transverse field Ising model using small neural networks. They emphasize that the ability to transfer policies from small samples to larger systems is useful insofar as the larger models can be implemented as a special-purpose experiment but solved using a noisier general-purpose quantum computer.

In both cases there is future work to be done extending the approaches to more problems and comparing reinforcement learning algorithms on QAOA specific tasks.  In fact, the literature seems so focused on PPO algorithms that a comparison may be in order.  Also, how implementation-level details within the PPO algorithm affect QAOA performance remains unclear.  Altogether, we expect the intersection of RL and QAOA to have plenty of room for research in both directions.

\subsection{Genetic Algorithms}

Genetic algorithms, introduced by Holland~\cite{genetic_first}, draw inspiration from genetics to find global optima in complex search spaces.  The idea is to generate populations of feasible solutions, subject them to something like natural selection, crossover candidates reproductively, then mutate the results.  Repeating this process draws the population into maxima while reserving the possibility to jump out of local traps~\cite{genetic_survey}.  In this way genetic algorithms have similar advantages as reinforcement learning in QAOA, but find more interesting applications in the compilation of quantum circuits.  

In a similar vein as the QCC-ACO of Section IV-A, Rasconi and Oddi use genetic algorithms to compile quantum circuits with minimum makespan, optimizing for the limitations of NISQ devices~\cite{geneticqcc}.  Their algorithm relies on an encoding whereby candidates' individual genes select quantum gates to be added to the circuit. They test their results on QAOA instances for the  MaxCut problem and find encouraging results, beating benchmark values in many cases. In the future, this approach ought to be tested on more problems.

A similar approach was taken by Arufe et al., who also benchmarked their decomposition-based genetic algorithm with QAOA~\cite{geneticcompile}.  Their algorithm ``outperformed the best current method on the largest instances and provided new best solutions to most of them'', supporting  genetic algorithms as a path forward for reliable near-term quantum circuit compilation.  Future genetic algorithms could push this field further, perhaps for problems with multiple objectives.

\subsection{Simulated Annealing}

Simulated annealing, a metallurgy-inspired method for combinatorial optimization, proceeds by first ``melting'' a population of solutions, then ``cooling'' until it ``freezes''~\cite{sim_anneal}.  These ``high'' temperatures allow for larger changes in the candidate state, and as the temperature goes down only smaller changes are allowed.  Then the ``freezing'' is achieved by satisfying some convergence criterion, completing the process.

This method has been compared with quantum annealing, which can be viewed as a variant of the same technique.  Quantum annealing, proposed by Kadowaki and Nishimori~\cite{quantumannealing_sa}, replaces the thermal effect-driven transitions of simulated annealing with  quantum tunneling. They show that quantum annealing converges to the optimal solution of the transverse Ising model obeying the time-dependent Schr\"odinger equation with higher probability than simulated annealing. Similarly, Farhi et al.~\cite{farhi_sa} give examples both of combinatorial optimization problems where the techniques perform similarly and where quantum annealing works better.

More recently, quantum annealing has been compared to simulated annealing in the contexts of wireless network scheduling~\cite{wirelessqa}.  This work should be compared with the application of QAOA by Choi et al.~\cite{networkqaoa} since QAOA is itself an approximation of quantum annealing.  Putting the pieces together, therefore, QAOA is closely related to simulated annealing.   Borle et al.~\cite{qaoa_sa} show that simulated annealing sometimes beats QAOA for a small number of angles, although QAOA may gain an advantage in the limit as the number of angles goes to infinity.  This may point to problems on NISQ devices, calling for more investigation.

\subsection{Summary:  QAOA and Computational Intelligence}

How QAOA fits into the overall landscape of computational intelligence is a topic that requires more research.  As a distant relative of simulated annealing, it connects with these techniques in various ways.  We have seen that computational intelligence can be used to simplify the search space for QAOA instances~\cite{acoqaoa}, and to compile the QAOA circuits themselves~\cite{qccaco,geneticqcc,geneticcompile}.  They can also be used to train an optimization algorithm for the QAOA VQA~\cite{reinforcementqaoa,qaoareinforcement}.

%\begin{figure*}[t]
%    \centering
%    \includegraphics[scale=0.5]{circuitn2.png}
%    \caption{A circuit diagram in Qiskit~\cite{Qiskit} of the TSP phase separator $U_P(1)$ for a symmetric graph with two cities and $d(u,v)=1$.  The quantum ZZ-gate $R_{ZZ}(\theta)=_{df}\exp(-i\frac{\theta}{2}Z\otimes Z)$ is a useful tool for implementation taken from the Qiskit circuit library.  This overall design is justified by how exponents distribute over sums of commutative operators.  There is no measurement in the  phase separation, so this circuit proceeds to the right into the mixing stage.}
%\end{figure*}

But QAOA is also a competitor to these classical approaches, and is designed to deal with large, constrained, and non-convex search spaces.  QAOA may be more efficient than the classical approaches for problems with search space containing many local minima~\cite{gradient_saddle}, although care must be taken to protect the classical optimization loop of the VQA from the impact of local minima.  QAOA is a very robust metaheuristic for discrete problems, whereas approaches such as ACO work best when the solution space has a certain structure (i.e., for the ants to traverse).  However, QAOA becomes more complicated for continuous problems. 

Also, QAOA may yet achieve approximations outperforming classical approaches. With more reliable quantum computers coming in the medium term as qubits become more robust and error-correcting improves, QAOA will be able to operate with a larger number $p$ of layers.  This may lead us to closer approximation values to the optimal, as the ansatz has a theoretical convergence to a correct solution in the limit of large $p$.  However, even in the cases where QAOA has no theoretical guarantee, it may be used as a heuristic to find better solutions than classical heuristics. The hope is that, in some cases, the resulting solutions may be provably better than that obtained by classical counterparts. For example, Farhi et al. proved a lower bound of 0.6924 for the approximation ratio of MaxCut instances on $3$-regular graphs for Farhi's QAOA with one level ($p=1$), and better bounds are known for $p>1$ on high-girth $3$-regular graphs~\cite{wurtz2021,caha2022}.  Although these do not outperform the best known classical algorithms, they are indicative of the type of results that may be possible.

\section{Illustrative QAOA Implementations}
\label{sec5}

\begin{figure*}
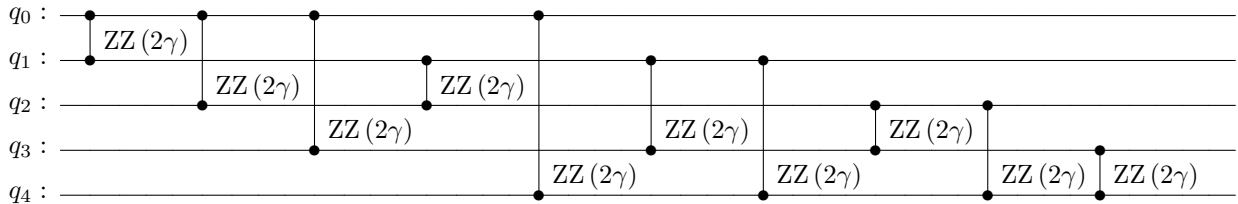
\label{cut_separator}
    \input maxcut_phase_separator
    \caption{The MaxCut phase separator $U_C(\gamma)$ for the complete graph on $5$ vertices.  Any other graph on $5$ vertices can be constructed from this by removing $R_{ZZ}$ interactions (which corresponds to removing edges).  The angles given to the gates are $2\gamma$ for technical reasons from Qiskit: since $R_{ZZ}(\theta)=_\text{df}\exp(-i\frac{\theta}{2}Z\otimes Z)$, an input of $\theta=2\gamma$ produces the desired gate.  This design is justified by how exponents distribute over sums of commutative operators, and the diagram uses the symmetry of $R_{ZZ}$.  This can be modified for the weighted case by multiplying the angle $2\gamma$ by $\alpha_{u,v}$ for each interaction between $u$ and $v$.}
\end{figure*}

In this section, we demonstrate how the QAOA can be used as a computational intelligence tool in its own right.  First, following the history of the field we describe a QAOA implementation for the MaxCut problem.  Then, illustrating the expanded power of the operator ansatz generalization, we modify this for the maximum bisection (MaxBis) problem. MaxBis is the problem of finding the maximum cut with equally sized partitions.  Here we follow Farhi et al.~\cite{qaoafarhi} for MaxCut and Golden et al.~\cite{golden2023numerical} for MaxBis.  Next we look at the   traveling salesperson problem (TSP), which is a central problem in computational intelligence used as a benchmark for optimization algorithms~\cite{tspbenchmark}. We use it as an example to illustrate the process of parameter selection and implementation.  Application of the QAOA to the TSP was studied by Hadfield et al.~\cite{hadfield2019}, which we follow closely in this subsection, and B\"{a}rtschi and Eidenbenz~\cite{baertschi2020grover}.  The TSP also provides a good jumping off point for exploring the connections between QAOA and CI since Ghimire et al.~\cite{acoqaoa} apply parallel ant-colony optimization to TSP QAOA.  For more information about this work, see Sec.~\ref{sec:aco}.

We print quantum circuit figures as Qiskit circuit diagrams~\cite{Qiskit} to help readers unfamiliar with quantum computing toolkits verify their code.  The horizontal lines represent qubits, initialized to the $\ket{0}$ state.  Geometric figures connecting qubits represent quantum gates, which are applied in series from left to right until control reaches the end of the diagram.  These gates in this section commute, allowing for the implementation trick $\exp(A+B)=\exp(A)\exp(B)$ given by the commutative case of Trotterization---see Sec.~\ref{sec:qaoalgorithm} for more information.

\subsection{Two Graph Cut Maximization Problems}
 \label{sec:twocutprobs}
Recall that a \textit{cut} on a graph $G=(V, E)$ is a partition of $V$ into two complementary subsets $V=V_1\sqcup V_2$.  The cost of a cut is the number of edges connecting the two subsets
\[C(V_1,V_2)=\#\{(v,w)\in V^2:v\in V_1, w\in V_2\},\]

\noindent which we want to maximize.  This problem is NP-complete and has been approached with computational intelligence techniques such as ant colony optimization~\cite{GAO20081173} and genetic algorithms~\cite{kim2001hybrid}. Throughout this section, we set $n=|V|$ for convenience.  In Sec.~\ref{maxcut}, we present the original QAOA algorithm for unconstrained maxcut, then in Sec.~\ref{maxbis}, we use the operator ansatz framework to solve the constrained variant called maximum bisection for graphs with an even number of nodes, which restricts the search space to graph bisections, which correspond to Hamming weight $n/2$ bitstrings.
\subsubsection{The Unconstrained Maximum Cut Problem}\label{maxcut}

Following Farhi et al.~\cite{qaoafarhi}, we use the transverse field mixer 
\[U_M(\beta)=\exp\left(-i\beta\sum_{j=0}^{n-1} X_j\right)=\prod_{j=0}^{n-1}\exp\Big(-i\beta X_j\Big),\text{ and}\]

\[U_C(\gamma)=\exp\left(-i\gamma\sum_{(v,w)\in E}Z_vZ_w\right)=\prod_{(u,v)
\in E}\exp\Big(-i\gamma Z_vZ_w\Big)\]

\noindent the phase separator for  $g(z)=2C(z)+2|E|$, which is maximized by the same bitstrings as $C$ is.  Taking the initial state to be the plus state completes the design of this QAOA instance for unconstrained maximum cut (Farhi's QAOA).

\subsubsection{The Maximum Bisection Problem}\label{maxbis}

The maximum bisection problem for graphs with an even number of nodes modifies MaxCut by restricting the search space to graph bisections, which are partitions into complementary sets of equal size.  In terms of the problem encoding, this corresponds to the set $S$ of Hamming weight $n/2$ strings of length $n$.  As this has the same objective function as MaxCut, the phase separator stays the same.  But we need to start and stay in the feasible subspace between QAOA rounds, so change the plus state to the superposition of Hamming weight $n/2$ strings of length $n$.  Such a superposition of equal Hamming-weight strings is called a \textit{Dicke state}.  There is a linear-depth, ancilla-free circuit to prepare Dicke states%given in B\"artschi and Eidenbenz
 \cite{baertschi2019deterministic}. This approach improves on the strategy of preparing any particular Hamming weight $n/2$ string proposed by Hadfield et al.~\cite{hadfield2019} by more closely preserving the original superposition method from Farhi's QAOA~\cite{cook2020vertexcover}. To be sure that mixing preserves the feasible subspace, one can use XY-interaction clique and ring mixers, which are called \textit{$XY$-mixers}~\cite{wang2020} and given respectively by
\[U_M^\text{clique}(\beta)=\exp\left(-i\beta \sum_{j>i}X_iX_j+Y_iY_j\right),\text{ and}\]
\[U_M^\text{ring}(\beta)=\exp\left(-i\beta \sum_{j\equiv i+1}X_iX_j+Y_iY_j\right)\]

\noindent where $\equiv$ represents equivalence modulo $|V|-1$.  These mixers only reach states in the feasible subspace of Hamming weight $k$ strings; the only other mixer with this feature in the literature is the Grover mixer, which uses the Dicke state preparation circuit mentioned above.  This mixer is given by
\[U_M^\text{grover}(\beta)=\exp\Big(-i\beta\ket{S}\bra{S}\Big),\]

\noindent where the state $\ket{S}$ is the superposition of feasible subspace basis states.  An implementation of the MaxCut phase separator is shown in Fig.~8, and the ring mixer in Fig.~9. For an implementation of the Grover mixer, see B\"artschi and Eidenbenz~\cite{baertschi2020grover}.

\begin{figure*}
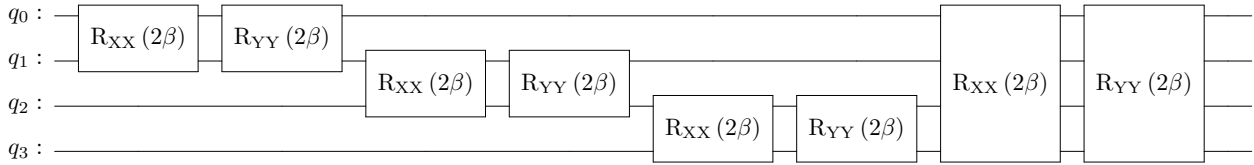
\label{ring_mixer}
    \centering\scalebox{0.9}{
        \input maxcut_ring_mixer
    }
    \caption{The ring mixer $U_M^\text{ring}(\beta)$ for any graph on $4$ vertices.  This mixer preserves the Hamming weight of computational basis states, thereby mixing the feasible subspace for maximum bisection.  The circuit comes with the added bonus of using only nearest-neighbor connections for efficient imlpementations on superconducting qubit quantum computing platforms.  By comparison, on such platforms the clique mixer requires SWAP instructions to run.}
\end{figure*}

\subsection{Further Extensions for MaxCut Problems}

It should be clear from this section how to extend these methods to more sophisticated problems.  For instance, both problems from Sec.~\ref{sec:twocutprobs} can be extended to weighted counterparts.  This corresponds to changing the cost function, but not the feasible subspace, so the mixer can stay the same while the phase separator changes.  We end up with the mixer
\[U_C(\gamma)=\prod_{(u,v)\in E}\exp\Big(-i\gamma \alpha_{v,w}Z_vZ_w\Big),\]
\noindent where the symbol $\alpha_{v,w}$ represents the weight of edge $(v,w)$.  Any other constant partition size restriction (e.g. one partition must have size $k$ and the other must have $n-k$) can be treated exactly as with maximum bisection.  In this case, all you need to change is the initial state, to the correctly-weighted Dicke state.  Hadfield et al.~\cite{hadfield2019} additionally describe a phase separator for a directed variant, Max-Directed-Cut, which can itself be extended to a weighted version just like described above. 

\subsection{The Travelling Salesperson Problem}
Recall that the TSP is the problem of finding a minimum-cost Hamiltonian cycle in a complete weighted digraph.  In the case where edges have the same cost in both directions, we call the TSP \textit{symmetric} and \textit{asymmetric} otherwise.  In this section we describe a QAOA solution for TSP which works on both symmetric and asymmetric instances.  The analogous problem for proper subdigraphs of the complete digraph can be approached by adding in sufficiently long edges.  

Let $[n]:=\{1,2,\ldots, n\}$.  We represent complete weighted digraphs on $n$ vertices as pairs $(V,d)$ where $V=[n]$ is the set of nodes, and $d:V^2\to \mathbf{R}$ is the cost function weighting the edges of the graph.  An instance of the traveling salesperson problem is a particular complete weighted digraph.  

The first question to ask is how to represent the solution space of this problem.  Hamiltonian cycles are equivalent to permutations of $[n]$, so we can encode solutions using $n^2$ qubits using a one-hot encoding.  For example, take the permutation $231$ corresponding to the cycle $2312$.  We can represent this by the \textit{one-hot encoding}~\cite{fuchs2021}
\[010+_{concat}001+_{concat}100=010001100,\]
where each city in the sequence is represented by a single bit.

Now what should be chooen as quantum parameters for this?  Take the convention that $\sigma_{n+1}=\sigma_1$ and the family of phase separating operators given by
\[U_P(\gamma)=\exp\left(-i\gamma\sum_{i=1}^n\sum_{u=1}^n\sum_{v=1}^n d(u,v) Z_{i,u}Z_{i+1,v}\right)\]

\noindent where $Z_{i,u}$ is the Pauli-Z gate that returns $1$ if city $u$ is in position $i$ and $-1$ otherwise.  The Hamiltonian used to construct this family of operators corresponds to the modified phase function $g(\sigma)=4C(\sigma)+(n-4)\sum_{u,v}d(u,v)$, where $C(\sigma)$ is the total cost of the cycle $\sigma$.  See Fig. 10 for the computation---we arrived at a different expression for $g(\sigma)$ compared to~\cite{hadfield2019}. This $C$ is what we are trying to minimize, so optimizing $g$ suffices since it is a linear formula in $C$ with positive slope.  Use the \textit{simultaneous order swap mixers}
\[U_M(\beta)=\exp\left (-i\beta\sum_{i=1}^n H_{PS}(i)\right)\]

\noindent where $H_{PS}(i)$ is the \textit{adjacent value-independent ordering swap partial mixing Hamiltonian}  swapping the cities at positions $i$ and $i+1$.   This $H_{PS}$ can be written in terms of \textit{individual adjacent value-selective ordering partial mixers}~\cite{hadfield2019}
\[H_{PS}(i,u,v)=S^+_{i, u}S^+_{i+1, v}S^-_{i,v}S^-_{i+1,u}+S^-_{i,u}S^-_{i+1,v}S^+_{i,v}S^+_{i+1,u}\]

\noindent swapping cities $u$ and $v$ only if they are in positions $i$ and $i+1$ (regardless of order).  Here $S^+_{j,u}=X+iY=\ket{1}\bra{0}$ and $S^{-}_{j,u}=X-iY=\ket{0}\bra{1}$, acting on the qubit encoding the fact that city $u$ is in position $j$.  Of course, using similar partial mixers we can construct more sophisticated mixers such as the \textit{color-parity ordering swap mixer} of Hadfield el al., which allows for a shallower circuit; but simultaneous order swapping is a rather straightforward choice fit for an introduction.

Since every permutation can be reached by performing some number of adjacent city swaps, this mixer satisfies the transition condition (equation 3).  It remains to choose an initial state.  
One way is to choose a particular permutation, for simplicity the identity permutation $e_{S_n}$.  The one-hot encoding of this as a quantum state is given by
\[\ket{e_{S_n}}=100\cdots 00+_{concat}010\cdots 00+_{concat}000\cdots01.\]

\begin{figure*}[t]
    \centering
    \includegraphics[scale=.375]{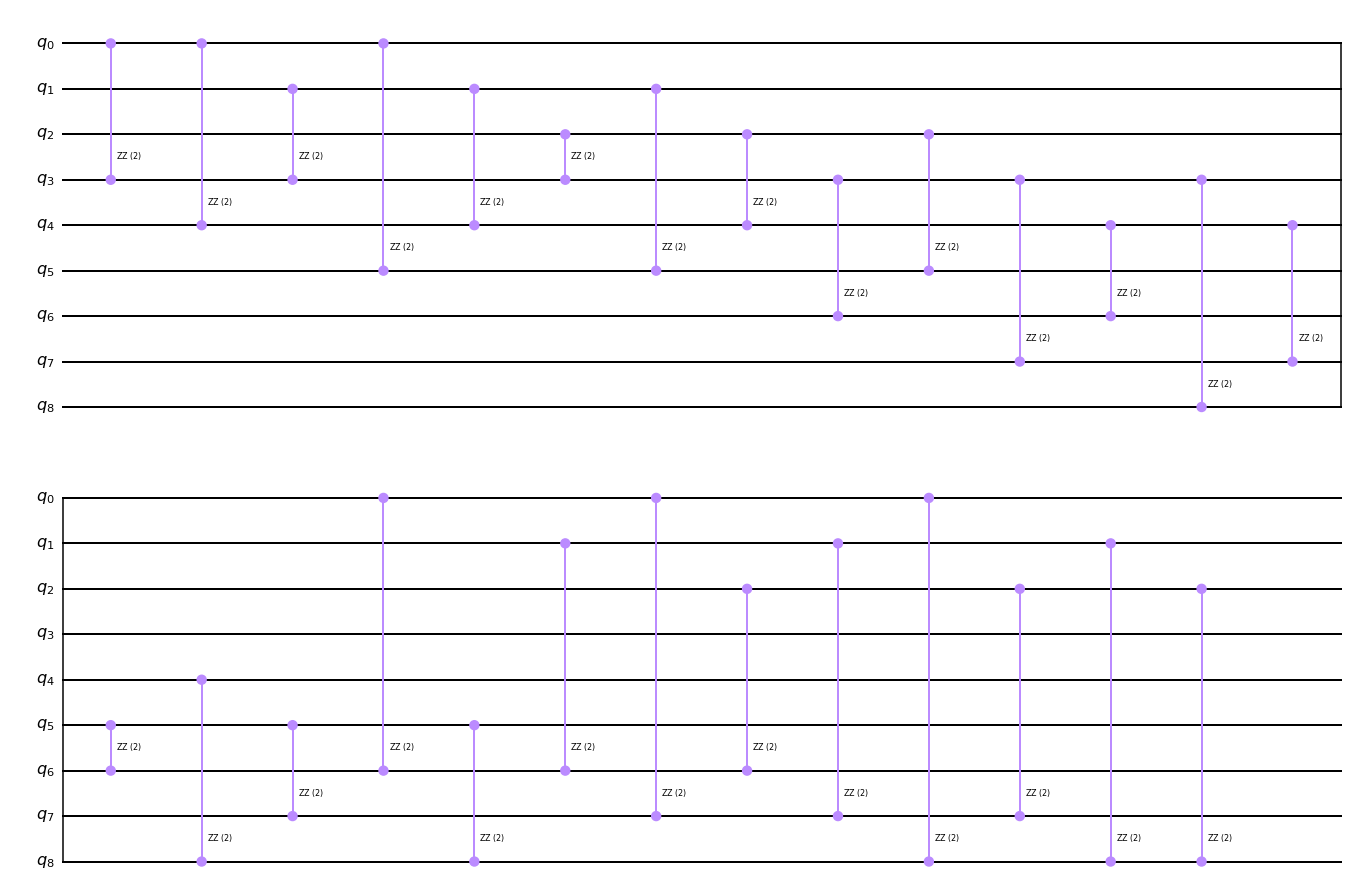}
    \caption{The TSP phase separator $U_P(1)$ for a symmetric graph with three cities and $d(u,v)=1$.  The quantum ZZ-gate $R_{ZZ}(\theta)=_{df}\exp(-i\frac{\theta}{2}Z\otimes Z)$ is a useful tool for implementation taken from the Qiskit circuit library.  In the asymmetric TSP with three cities, there are two distinct feasible solutions in the asymmetric case (making this the first interesting case).  Because the number of qubits needed for this approach scales is $n^2$ and the number of gates is $n^3$, it is uneconomical to print larger examples.  However, bigger circuits are constructed in an analogous way that these smaller examples should make clear.}
\end{figure*}

\noindent We can construct this state using just $n$ Pauli-X gates by
\[U_{\ket{e_{S_n}}}=X_{1,1}+X_{2,2}+\cdots +X_{n,n},\]

\noindent easily generalizing to an arbitrary permutation.  This quantum ansatz for the TSP can be constructed with a polynomial number of gates for the phase separator and mixer,  with $\mathcal{O}(n^3)$ gates specifically. Due to the triviality of the state preparation operator, this is the total complexity.

A different starting state is the full superposition of all cycles/permutations. This has been shown to be possible with $\mathcal{O}(n^3)$ gates in $\mathcal{O}(n^2)$ depth, even 
on a limited linear nearest neighborhood hardware connectivity~\cite{baertschi2020grover}. The main idea for their circuit is to replace fixed one-hot encodings for sequence entries by $W$-states (superpositions of one-hot encodings), with a bit-mask controlling that a city does not appear multiple times. Importantly, as mentioned before, such a state preparation gives rise to a Grover mixer of the same complexity, which can be explored as a competitor to the previously mentioned mixers.

All of these variants are fairly easy to implement, but come with some significant drawbacks.  For one thing, the one-hot encoding of the permutation puts large graphs out of reach on NISQ devices.  Indeed, current devices with $127$ qubits can only handle graphs up to size $n=11$.  The depth of the circuit, being $\mathcal{O}(n^2)$, is also problematic for large $n$.  Therefore we either need serious simplifications for this ansatz to be workable on NISQ devices, or future fault-tolerant hardware to run it for larger instances. However, it runs on toy examples and it nicely exemplifies the QAOA design and implementation processes.  For an illustrative example of the implementation of the TSP phase separator  for a graph with $3$ vertices (hence a $9$-qubit encoding), see Fig.~9. 

As a final note, this and other constrained problems can be approached as unconstrained problems using penalty terms keeping us in the feasible subspace~\cite{qaoa_penalties}, but this will often cost more in compilation complexity and circuit depth.

\section{Conclusion}
\label{sec6}

In this introductory tutorial paper, we presented a broad overview of the quantum alternating operator ansatz  and how it serves as a metaheuristic for combinatorial optimization problems.  We introduced quantum computing for a general audience, and explained the derivation and history of QAOA.  We categorized its hybrid parameters by their quantum and classical nature, describing a general framework for designing QAOA instances for problems of interest in optimization.
\begin{figure*}
  \begin{align*}
  H_{TSP}\ket{z}&=\left(\sum_{i=1}^n \sum_{u=1}^n\sum_{v=1}^n d(u,v)Z_{i,u}Z_{i+1,v}\right)\ket{z} \\
  &=\sum\limits_{i=1}^n \Bigg(\underline{\underline{\underline{d(\sigma_i,\sigma_{i+1})}}}-\underline{\sum\limits_{u\neq \sigma_i}{d(u,\sigma_{i+1})}}-\underline{\sum\limits_{v\neq \sigma_{i+1}} d(\sigma_i,v)}+\underline{\underline{\sum\limits_{\substack{u\neq \sigma_i\\v\neq\sigma_{i+1}}}d(u,v)}}\Bigg)\ket{z}\\
  &=\left(\underline{\underline{\underline{C(\sigma)}}} + (\underline{\underline{n-2}}\ \underline{-2}) \sum\limits_{uv\not\subseteq \sigma} d(u,v)+ \underline{\underline{(n-1) \sum\limits_{uv\subseteq \sigma} d(u,v)}}\right)\ket{z}\\
  &=\underline{\underline{\underline{\sum\limits_{uv\subseteq \sigma}d(u,v)}}} + (n-4) \sum\limits_{uv\not\subseteq \sigma} d(u,v)+ (n-1) \sum\limits_{uv\subseteq \sigma} d(u,v)\Bigg)\ket{z} \\
  &=\Bigg(n \sum\limits_{uv\subseteq \sigma} d(u,v) + (n-4) \sum\limits_{uv\not\subseteq \sigma} d(u,v)\Bigg)\ket{z}\\
  &=\Bigg(4 C(\sigma)+(n-4)\sum\limits_{uv} d(u,v)\Bigg)\ket{z}=g(\sigma)\ket{z} 
  \end{align*}
  \caption{Computing the TSP phase function.  For each $(i,u,v)\in [n]^3$ we have that the monomial $d(u,v)Z_{i,u}Z_{i+1,v}$ equals $-d(u,v)$ if exactly one of (i) $u$ occurs at position $i$ or (ii) $v$ occurs at position $i+1$, or $d(u,v)$ otherwise -- i.e., if both or neither of (i) or (ii) apply.  Using this combinatorial interpretation, we can compute the Hamiltonian on a given bitstring $z$ corresponding to a permutation $\sigma$ with $\sigma_{n+1}=_{df}\sigma_1$.  The underlining is intended to help track what is transforming into what throughout the calculation, and here the nonstandard notation $uv\subseteq \sigma$ means that $uv$ is a substring of the cycle $\sigma+_{concat}\sigma_{n+1}$.}
  \end{figure*}
Further, we connected our work to computational intelligence (CI), discussing how well-known CI approaches have been applied to QAOA and ideas for how this work can be extended in the future.  
We also explain how QAOA can serve alongside these techniques, given improvements in quantum computers.  To demonstrate how to implement QAOA, we discuss three optimization problems: maximum cut, maximum bisection, and the traveling salesperson problem.  We give a brief tutorial on how to implement QAOA instances for these problems and their weighted counterparts, mostly focusing on the quantum parameters of phase separators, mixers, and initial states (but with necessary consideration also given to problem encodings).  The other parameters (classical optimization method and number of angles) are less clearly derived from the problem, and more often determined by trial and error.

%and extensions of this algorithm, and how it relates to ant colony optimization, reinforcement learning, genetic algorithms, and simulated annealing.  We present a bifurcation of the parameter space for QAOA techniques into classical and quantum categories, together with references to studies on examples of particular choices.
%We illustrate the QAOA algorithm design process with a discussion of the ubiquitous and NP-hard traveling salesperson problem, giving details from our implementation of the TSP phase separator proposed by Hadfield et al.~\cite{hadfield2019} in Qiskit, with circuit diagrams to illustrate the technique for small graphs.  %We hope this example of the procedure can help others to map more specific problems of interest in engineering onto the QAOA ansatz, which is an important next step in research.

%These and related ideas should lead to more work in the intersection between such techniques, and help to expand both fields as hybrid algorithms involving both topics are an exciting topic of ongoing study.  
As noisy intermediate scale quantum (NISQ) devices improve on reliability and error correction, QAOA on quantum computers may give the best approximations due to their theoretical potential. 
%This may eclipse classical techniques for certain problems. 
As noted, even in cases where convergence is difficult, QAOA can be used to find good solutions, which for certain problems may be provably better than those obtained with classical techniques.  We expect more work to appear connecting QAOA to Computational Intelligence.

\section{Acknowledgements}

This work was supported in part by the U.S. Department of Energy through the Los Alamos National Laboratory. Los Alamos National Laboratory is operated by Triad National Security, LLC, for the National Nuclear Security Administration of U.S. Department of Energy (Contract No. 89233218CNA000001).  Research presented in this article was supported by the Laboratory Directed Research and Development program of Los Alamos National Laboratory under project number 20230049DR. Los Alamos National Laboratory Publication Number: LA-UR-23-20638. \textcolor{black}{This work was also sponsored by the US Department of Energy  under grant \#DE-SC0023392, and US NSF awards \#2148358 and \#1914635.}

\bibliography{bibtex}
\bibliographystyle{ieeetr}
% was ieeetr
\end{document}

%% file: maxcut_phase_separator.tex
\scalebox{1.0}{
\Qcircuit @C=1.0em @R=0.8em @!R { \\
	 	\nghost{{q}_{0} :  } & \lstick{{q}_{0} :  } & \ctrl{1} & \dstick{\hspace{2.0em}\mathrm{ZZ}\,(\mathrm{2{\ensuremath{\gamma}}})} \qw & \qw & \qw & \ctrl{2} & \qw & \qw & \qw & \ctrl{3} & \qw & \qw & \qw & \qw & \qw & \qw & \qw & \ctrl{4} & \qw & \qw & \qw & \qw & \qw & \qw & \qw & \qw & \qw & \qw & \qw & \qw & \qw & \qw & \qw & \qw & \qw & \qw & \qw & \qw & \qw & \qw & \qw & \qw & \qw\\
	 	\nghost{{q}_{1} :  } & \lstick{{q}_{1} :  } & \control \qw & \qw & \qw & \qw & \qw & \dstick{\hspace{2.0em}\mathrm{ZZ}\,(\mathrm{2{\ensuremath{\gamma}}})} \qw & \qw & \qw & \qw & \qw & \qw & \qw & \ctrl{1} & \dstick{\hspace{2.0em}\mathrm{ZZ}\,(\mathrm{2{\ensuremath{\gamma}}})} \qw & \qw & \qw & \qw & \qw & \qw & \qw & \ctrl{2} & \qw & \qw & \qw & \ctrl{3} & \qw & \qw & \qw & \qw & \qw & \qw & \qw & \qw & \qw & \qw & \qw & \qw & \qw & \qw & \qw & \qw & \qw\\
	 	\nghost{{q}_{2} :  } & \lstick{{q}_{2} :  } & \qw & \qw & \qw & \qw & \control \qw & \qw & \qw & \qw & \qw & \dstick{\hspace{2.0em}\mathrm{ZZ}\,(\mathrm{2{\ensuremath{\gamma}}})} \qw & \qw & \qw & \control \qw & \qw & \qw & \qw & \qw & \qw & \qw & \qw & \qw & \dstick{\hspace{2.0em}\mathrm{ZZ}\,(\mathrm{2{\ensuremath{\gamma}}})} \qw & \qw & \qw & \qw & \qw & \qw & \qw & \ctrl{1} & \dstick{\hspace{2.0em}\mathrm{ZZ}\,(\mathrm{2{\ensuremath{\gamma}}})} \qw & \qw & \qw & \ctrl{2} & \qw & \qw & \qw & \qw & \qw & \qw & \qw & \qw & \qw\\
	 	\nghost{{q}_{3} :  } & \lstick{{q}_{3} :  } & \qw & \qw & \qw & \qw & \qw & \qw & \qw & \qw & \control \qw & \qw & \qw & \qw & \qw & \qw & \qw & \qw & \qw & \dstick{\hspace{2.0em}\mathrm{ZZ}\,(\mathrm{2{\ensuremath{\gamma}}})} \qw & \qw & \qw & \control \qw & \qw & \qw & \qw & \qw & \dstick{\hspace{2.0em}\mathrm{ZZ}\,(\mathrm{2{\ensuremath{\gamma}}})} \qw & \qw & \qw & \control \qw & \qw & \qw & \qw & \qw & \dstick{\hspace{2.0em}\mathrm{ZZ}\,(\mathrm{2{\ensuremath{\gamma}}})} \qw & \qw & \qw & \ctrl{1} & \dstick{\hspace{2.0em}\mathrm{ZZ}\,(\mathrm{2{\ensuremath{\gamma}}})} \qw & \qw & \qw & \qw & \qw\\
	 	\nghost{{q}_{4} :  } & \lstick{{q}_{4} :  } & \qw & \qw & \qw & \qw & \qw & \qw & \qw & \qw & \qw & \qw & \qw & \qw & \qw & \qw & \qw & \qw & \control \qw & \qw & \qw & \qw & \qw & \qw & \qw & \qw & \control \qw & \qw & \qw & \qw & \qw & \qw & \qw & \qw & \control \qw & \qw & \qw & \qw & \control \qw & \qw & \qw & \qw & \qw & \qw\\
\\ }}

%% file: maxcut_ring_mixer.tex
\scalebox{1.0}{
\Qcircuit @C=1.0em @R=1.0em @!R { \\
	 	\nghost{{q}_{0} :  } & \lstick{{q}_{0} :  } & \multigate{1}{\mathrm{R_{XX}}\,(\mathrm{2{\ensuremath{\beta}}})} & \multigate{1}{\mathrm{R_{YY}}\,(\mathrm{2{\ensuremath{\beta}}})} & \qw & \qw & \qw & \qw & \multigate{3}{\mathrm{R_{XX}}\,(\mathrm{2{\ensuremath{\beta}}})} & \multigate{3}{\mathrm{R_{YY}}\,(\mathrm{2{\ensuremath{\beta}}})} & \qw & \qw\\
	 	\nghost{{q}_{1} :  } & \lstick{{q}_{1} :  } & \ghost{\mathrm{R_{XX}}\,(\mathrm{2{\ensuremath{\beta}}})} & \ghost{\mathrm{R_{YY}}\,(\mathrm{2{\ensuremath{\beta}}})} & \multigate{1}{\mathrm{R_{XX}}\,(\mathrm{2{\ensuremath{\beta}}})} & \multigate{1}{\mathrm{R_{YY}}\,(\mathrm{2{\ensuremath{\beta}}})} & \qw & \qw & \ghost{\mathrm{R_{XX}}\,(\mathrm{2{\ensuremath{\beta}}})} & \ghost{\mathrm{R_{YY}}\,(\mathrm{2{\ensuremath{\beta}}})} & \qw & \qw\\
	 	\nghost{{q}_{2} :  } & \lstick{{q}_{2} :  } & \qw & \qw & \ghost{\mathrm{R_{XX}}\,(\mathrm{2{\ensuremath{\beta}}})} & \ghost{\mathrm{R_{YY}}\,(\mathrm{2{\ensuremath{\beta}}})} & \multigate{1}{\mathrm{R_{XX}}\,(\mathrm{2{\ensuremath{\beta}}})} & \multigate{1}{\mathrm{R_{YY}}\,(\mathrm{2{\ensuremath{\beta}}})} & \ghost{\mathrm{R_{XX}}\,(\mathrm{2{\ensuremath{\beta}}})} & \ghost{\mathrm{R_{YY}}\,(\mathrm{2{\ensuremath{\beta}}})} & \qw & \qw\\
	 	\nghost{{q}_{3} :  } & \lstick{{q}_{3} :  } & \qw & \qw & \qw & \qw & \ghost{\mathrm{R_{XX}}\,(\mathrm{2{\ensuremath{\beta}}})} & \ghost{\mathrm{R_{YY}}\,(\mathrm{2{\ensuremath{\beta}}})} & \ghost{\mathrm{R_{XX}}\,(\mathrm{2{\ensuremath{\beta}}})} & \ghost{\mathrm{R_{YY}}\,(\mathrm{2{\ensuremath{\beta}}})} & \qw & \qw\\
\\ }}